\documentclass[useAMS,usenatbib,usegraphicx]{mn2e}
\usepackage{times}
\usepackage{longtable} 
\usepackage{lscape}
\usepackage{color}
\usepackage{amssymb}

\def\cm3{$\rm cm^{-3}$}

\def\n0{$\rm n_{0}$}
\def\B0{$\rm B_{0}$}

\def\mc{$\mu$m}

\def\L12{L$_{12\mu m}$~}
\def\F12{F$_{12\mu m}$~}

\def\fe2{[Fe\,{\sc ii}]}
\def\s3{[S{\sc iii}]}
\def\h2{H$_{2}$}

\def\F{$F_{\lambda}$}

\def\pp{$\pm$}

\title[Stellar Populations in the NIR]{Probing the Circumnuclear Stellar Populations of Starburst Galaxies in the Near-infrared}

\author[N.Z. Dametto et al.]{N.Z. Dametto$^{1}$\thanks{E-mail:
natacha.zanon@ufrgs.br}, R. Riffel$^{1}$, M. G. Pastoriza$^{1}$, A. Rodr\'{\i}guez-Ardila$^{2}$\thanks{Visiting Astronomer at the Infrared Telescope Facility, which is operated by the University of 
Hawaii under Cooperative Agreement no. NCC 5-538 with the National Aeronautics and Space Administration, Office of Space Science, Planetary Astronomy Program.}, 
\newauthor{J. A. Hernandez-Jimenez$^{1}$, E. A. Carvalho$^{3,4}$.}
 \\$^{1}$Departamento de Astronomia, Universidade Federal do Rio Grande do Sul. Av. Bento Gon\c calves 9500, Porto Alegre, RS, Brasil.
\\$^{2}$Laborat\'{o}rio Nacional de Astrof\'{i}sica/MCTI - Rua dos Estados Unidos 154, Itajub\'{a}, MG, Brasil.
\\$^{3}$Universidade Federal de Itajub\'{a}. Rua Doutor Pereira Cabral, 1303, Pinheirinho, Itajub\'{a}, MG, Brasil.
\\$^{4}$Campus S\~ao Jo\~ao Del Rei, Instituto Federal do Sudeste de Minas. Rua Am\'{e}rico Davim Filho, s/n$^o$, S\~ao Jo\~ao Del Rei, MG, Brasil.}

\begin{document}

\date{Accepted . Received , in original form }
\pagerange{\pageref{firstpage}--\pageref{lastpage}} \pubyear{2014}

\maketitle

\label{firstpage}

\begin{abstract}
We employ the NASA Infrared Telescope Facility's near-infrared spectrograph SpeX at 0.8-2.4$\mu$m to investigate the spatial distribution 
of the stellar populations (SPs) in four well known Starburst galaxies: NGC\,34, NGC\,1614, NGC\,3310 and NGC\,7714. 
We use the {\sc starlight} code updated with the synthetic simple stellar populations models computed by \citet[ M05]{maraston05}.
Our main results are that the NIR light in the nuclear surroundings of the galaxies is dominated by young/intermediate age SPs 
($t \leq 2\times10^9$\,yr), summing from $\sim$40\% up to 100\% of the light contribution. In the nuclear aperture of two sources (NGC\,1614 and NGC\,3310)
we detected a predominant old SP component ($t > 2\times10^9$\,yr), while for NGC\,34 and NGC\,7714 the younger component prevails. 
Furthermore, we found evidence of a circumnuclear star formation ring-like structure and a secondary nucleus in NGC\,1614, in agreement with previous studies.
We also suggest that the merger/interaction experienced by three of the galaxies studied, NGC\,1614, NGC\,3310 and NGC\,7714 can explain
the lower metallicity values derived for the young SP component of these sources. In this scenario the fresh unprocessed metal poorer gas from 
the destroyed/interacting companion galaxy is driven to the centre of the galaxies and mixed with the central region gas, before star formation takes place.
In order to deepen our analysis, we performed the same procedure of SP synthesis using \citet[][ M11]{maraston11} EPS models.
Our results show that the newer and higher resolution M11 models tend to enhance the old/intermediate age SP contribution over the younger ages.

\end{abstract}


\section{Introduction}

A galaxy that is undergoing an intense star formation, usually in the central region (r $\lesssim$1\,kpc), 
is called a Starburst galaxy (SB). These objects present star-formation rates (SFRs) of 5-50M$_{\odot}$yr$^{-1}$ within a region
of 0.1-1\,kpc extent, exceeding the values found within a similar region in the Galactic center \citep[0.5M$_{\odot}$yr$^{-1}$, ][]{gusten89} or
in normal galaxies \citep[e.g. ][]{heck00}. Also, its spectrum is characterized by unusually 
bright emission lines, specially hydrogen and helium recombination lines.

It is widely known that what powers the starburst are massive stars that emit their primary radiation in the ultraviolet (UV) part of 
the spectrum \citep{heck00}, while the interstellar medium absorbs these radiation and re-radiates it at longer 
wavelengths. Massive stars evolves rapidly to red supergiants (RSGs) that emit most of their radiation  
in the near-infrared (NIR). These RSGs, therefore, are indicators of young stellar populations (SPs), 
providing means to identify recent starbursts in the NIR \citep{oliva95}. Thus, SBs are key sources to study 
the formation and evolution of massive stars in galaxies.

Moreover, the thermally pulsing asymptotic giant branch (TP-AGB) stellar phase has an important contribution in 
the NIR \citep{maraston98,maraston05,riffel08}, being enhanced in young to intermediate age SPs ($t < 2$ Gyr). 
With the new generation of the evolutionary 
population synthesis (EPS) models that include the empirical spectra of stars in the 
TP-AGB phase \citep[][hereafter M05]{maraston05}, it became possible to study in more 
details the SP of galaxies in the NIR. A further benefit with the inclusion of these empirical spectra 
of C and O-rich stars to the models \citep{lw00} was the detection 
of NIR characteristic absorption features from the CN band.
\citet{riffel07} for instance, detected the 1.1\mc\ CN band in the spectra of SBs and Seyfert galaxies.
In addition, \citet{martins13} detected the band at 1.1\mc\ and that at 1.4\mc\ of the same molecule.

Even though tracing star formation in the NIR is a difficult task, this spectral range is the most suitable one to 
unveil the SPs in highly obscured sources  \citep{origlia00}. With the improvement of infrared (IR) arrays, it became possible to obtain spectra
at moderate resolution on faint and extended sources \citep[][ hereafter R08]{riffel08} allowing a more detailed
study of SPs in the NIR spectral region. Also, the simultaneous wavelength coverage provided by these instruments via cross-dispersed
spectroscopy brings an improvement to the SP synthesis once it avoids the aperture and seeing effects that usually affects
the study in the {\it J. H} and {\it K} spectroscopy done in long-slit and single-band modes. It is worth mentioning that the 
James Webb Space Telescope (JWST) will be optimized for observations in the IR. Thus, it is important to test the
SP synthesis method in this spectral region.

In spite of the large amount of results about SPs gathered by means of NIR in active galactic nuclei (AGNs) or individual objects, few studies
have concentrated on the analysis of the SPs along the spatial directions at distances larger than a few hundred
of parsecs. It is thus necessary to develop proxies for the analysis of SPs along the radial direction and compare these
results with those found at other wavelengths. This allows us to find out complementary information that the SP synthesis
in the NIR is able to provide. To this purpose, we selected a sample composed by four well-known SBs in the local universe (NGC\,34, NGC\,1614, NGC\,3310, NGC\,7714), 
which were widely studied in the optical and NIR (see Sec.~\ref{previous}) and whose spatial extension and proximity allows the study of SPs at distances as large as 
several hundreds of parsecs. 

An important aspect regarding our sample is the fact that all four galaxies are merging systems. It is known that
mergers can trigger star formation \citep{mih96}, as well as create peculiar structures as tidal tails, bridges\citep[e.g.][]{toom72} and 
rings \citep[e.g.][]{lyn76}. An evidence of that is the presence of luminous H{\sc ii} regions in collisional rings \citep[e.g.][]{Fos77,mar95}, 
and in extended tidal tails \citep[e.g.][]{duc11}. In fact, interacting galaxies usually have strong IR emission
\citep[Luminous Infrared Galaxies - LIRGs and Ultra-Luminous 
Infrared Galaxies - ULIRGs\footnote{LIRGs are defined as $10^{11} \leq L/L_{\odot} < 10^{12}$ and ULIRGs as $L/L_{\odot} \geq 10^{12}$.},  e.g.][]{rieke80} and 
present different types of nuclear activity 
like nuclear starbursts, AGN as well as post-starburst activity. With this in mind, we also would like to emphasize the importance of the detailed study of 
star-forming interacting systems in the local universe using NIR spectral range. These results can provide further support to studies of high-z sources \citep[e.g.][]{pope13}, once almost all 
systems displayed intense bursts of star formation and were strongly interacting at the early universe \citep[z$\sim$1-5, ][]{carilli13}. 

Thus, aimed at studying the nuclear and off-nuclear SP in SBs and building a scenario of the star formation history (SFH) of these sources, 
we will analyze the spacial distribution of the SPs, in the NIR spectral range, of NGC\,34, NGC\,1614, 
NGC\,3310 and NGC\,7714, by means of stellar population fitting.

This paper is structured as follows. In Section 2.1 we present a summary of previous stellar population studies of our sample and in Section 2.2 the observations and data reduction.   
In Section 3 we describe the method used to perform the stellar population synthesis.
The results are presented and discussed in section 4 and the conclusions are left to Section 5. 
We assume $H_0$=75\,km\,s$^{-1}$Mpc$^{-1}$.


\section{The Data}\label{data}
\subsection{Sample and Previous Stellar Population Studies}\label{previous}

In this section we summarize the properties of the galaxies as well as the results of the previous SP studies of our sample.

{\bf NGC\,34:} 
at a distance of 78.4\,Mpc, this LIRG was classified as a merger system by \citet{vv59}. 
The nuclear activity classification of this galaxy is controversial. For example, \citet{mazz91,rrp06} classified this source
as a SB. Yet, according to \citet{gon99}, its optical nuclear spectrum is not only starburst-dominated, but also hosts a Seyfert\,2 (Sy\,2) nuclei. 
In fact, many studies have classified this source as a Sy\,2 \citep[e.g.][]{ver86, gol97a}, while
others have emphasized the apparent weakness of the [O{\sc iii}] $\lambda$5007 emission line relative to either H$\alpha$
or H$\beta$ and classified NGC\,34 as narrow-emission-line galaxy \citep[e.g.][]{oster83,gol97b}. 

Using optical images and spectroscopic observations \citet{sch07} proposed a scenario about the merging history of NGC\,34.
They suggested that two disk gas-rich galaxies of unequal mass (with an estimated mas ratio of 1/3 $\leq m/M \leq$ 2/3) merged,
yielding a galaxy-wide starburst. This starburst occurred first $\sim$600\,Myr (showing a peak over 100\,Myr ago) and, according to the authors, 
seems to have formed an extensive system of young globular clusters with ages in the range of 0.1 to 1.0\,Gyr. This work also reveals a young, blue stellar 
exponential disk that formed $\sim$400\,Myr ago. At present, the two merging galaxies' nuclei appear to have coalesced, the
starburst has shrunk to its current central ($\leq$ 1\,kpc) state and there is a strong gaseous outflow.

In addition, R08 analyzed the inner 230\,pc
of this source in the NIR and detected a young-intermediate-age SP with solar metallicity. Their
results are in agreement with the fact that this galaxy has a strong 1.1\mc\ CN absorption band in its spectrum
\citep{riffel07} characteristic of SPs with that age ($\sim$ 1\,Gyr, M05).

{\bf NGC\,1614:} 
at a distance of 63.7\,Mpc, NGC\,1614 is considered a suitable laboratory to study 
starbursts since it has a modest extinction and fairly face-on viewing geometry \citep{alo01}. This source is
cataloged as LIRG and shows a spectacular outer structure with tidal tails or plumes, suggesting 
that this morphology and its extreme IR luminosity results from the interaction/merger with at least other 
two galaxies \citep[e.g.][]{neff90, alo01}. Also, the highly asymmetric extended emission present around this source favors  
that this interaction scenario is still occurring \citep[e.g.][]{kot01}. 
\textit{Hubble Space Telescope/NIR} camera and multi-object spectrometer (NICMOS) observations reported by 
\citet{alo01} show deep CO stellar absorption, interpreted by them as tracers of a starburst nucleus of $\sim$45\,pc in diameter surrounded by a 
$\sim$600\,pc diameter ring of supergiant H{\sc ii} regions revealed in Pa$\alpha$ line emission. This ring is coincident with
a ring of radio continuum emission detected by \citet{olss10}, who conclude that the LINER-like activity 
observed in NGC\,1614 can be attributed to starburst activity, and not due to an AGN. \citet{alo01}
also states that the presence of a secondary nucleus can be interpreted as being fragments of the companion 
galaxy, smaller than NGC\,1614, which had long been destroyed. Thus this source was classified as an advanced merger system by \citet{doyon89,neff90}. 
Moreover, R08 detected a dominant 1\,Gyr old SP in the inner 154\,pc in agreement to the fact that the integrated spectra of both NGC\,34 and NGC\,1614
are very similar.

{\bf NGC\,3310:} 
it is a well known nearby SB (d=13.2\,Mpc) which presents a disturbed morphology \citep{elmg02}, suggesting that it recently undergone at least one significant merger.
\citet{balick81} were the first to propose these merging scenario, suggesting that star formation occurring in the past $10^8$ yr has been triggered by 
a collision with a dwarf galaxy. These authors also states that the well known `arrow' morphology \citep{wal67} on the western
side comprises the remnant of this dwarf galaxy. NGC\,3310 harbors a circumnuclear ring of about 8$\arcsec$ to 12$\arcsec$ in diameter \citep[720 - 1080\,pc][]{elmg02}
that has been studied over a wide range of wavelengths, such as x-rays \citep[e.g.][]{zezas98}, far-UV \citep[e.g.][]{smith96}, UV \citep[e.g.][]{meurer95}, 
optical \citep[e.g.][]{groth91, mgp93, balick81}, NIR \citep[e.g.][]{tega84,mgp93,elmg02}, IR \citep[e.g.][]{tega84} and radio \citep[e.g.][]{balick81}.
This ring is filled with giant H{\sc ii} regions \citep{mgp93}. 
Studying the inner 56\,pc of this source in the NIR spectral range, 
R08 detected an intense starburst activity, with four starbursts, one dominant at
1\,Gyr, which contributes with up to $\sim$30\% of the light in this spectral range.

{\bf NGC\,7714:} 
a SBc galaxy at 37.3\,Mpc classified as a prototypical SB by \citet{weed81}. This galaxy is in 
interaction/merger with its companion NGC\,7715, and such interaction could be responsible
for the star formation bursts \citep{kinney93}. In fact, \citet{lancon01} show the bridge between NGC\,7714 and its 
companion NGC\,7715 (see their Fig.\,2). This source is very well studied with large amount of data in 
the optical and NIR part of the spectrum \citep[e.g.][]{rosa95,rosa99,lancon01,brandl04}. 
For example, analyzing narrow band H$_{\alpha}$ imaging, \citet{rosa95} conclude that the star formation 
bursts in the inner 5$\arcsec$ (945\,pc) of the galaxy is a collection 
of smaller H{\sc ii} regions with ages between 4 to 5\,Myr. With very little silicate absorption and temperature of the hottest
dust component of 340\,K, NGC\,7714 is defined by \citet{brandl04} as the perfect template for young, unobscured starburst.
Moreover, analyzing the optical continuum and far-infrared (FIR) 
colors \citet{bern93} found that the star formation in this source is consistent with a continuous star 
formation rate during the past 20\,Myr. R08 detected three star formation bursts, one dominant at 1\,Gyr which 
contributes with up to $\sim$34\% of the light and two minor bursts with ages 30\,Myr ($\sim$10\%) and 50\,Myr ($\sim$13\%).

\subsection{Observations and data reduction}\label{observations}

\begin{figure*}
\begin{minipage}[b]{0.49\linewidth}
\includegraphics[width=\linewidth]{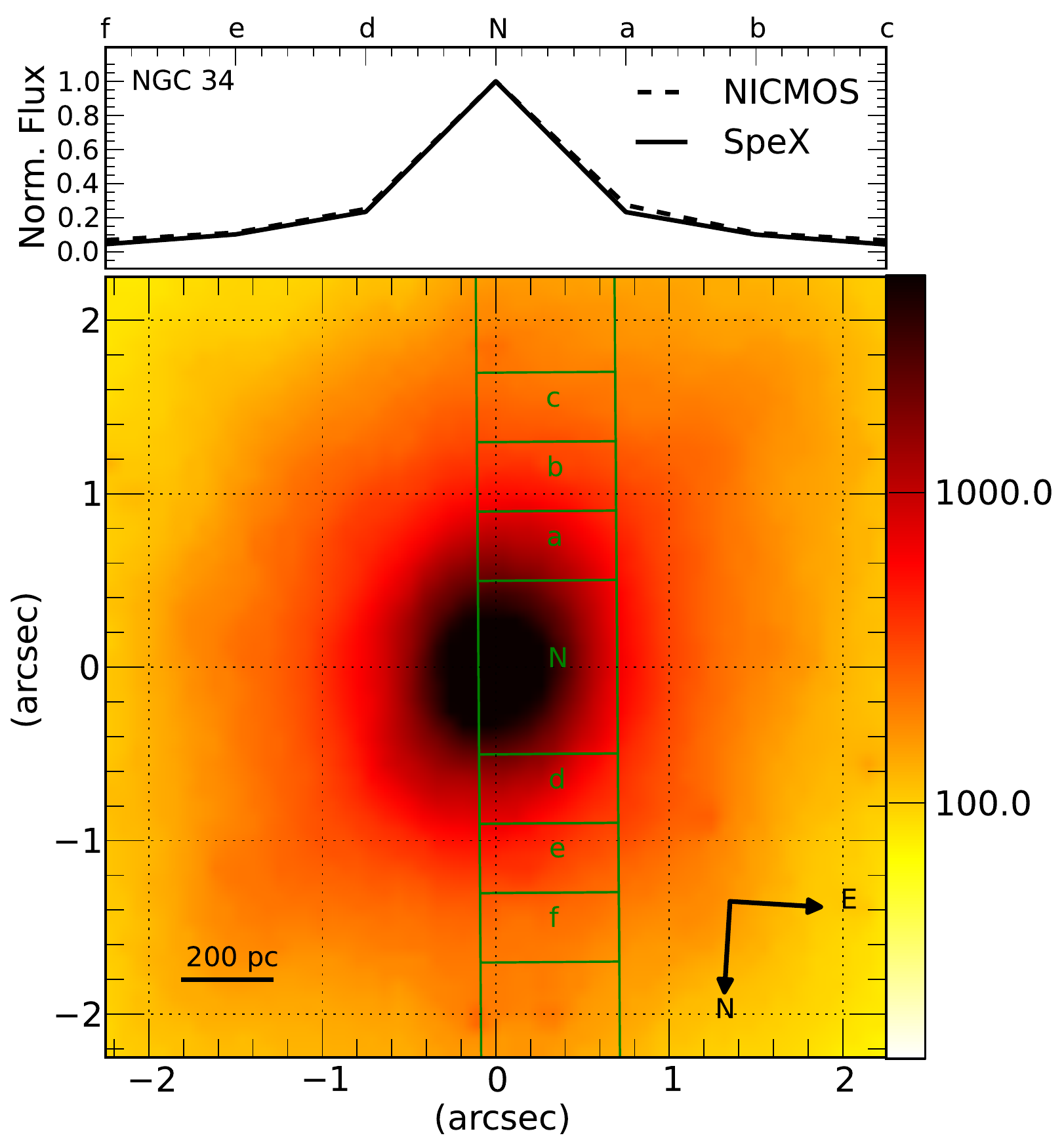}
\end{minipage} \hfill
\begin{minipage}[b]{0.49\linewidth}
\includegraphics[width=\linewidth]{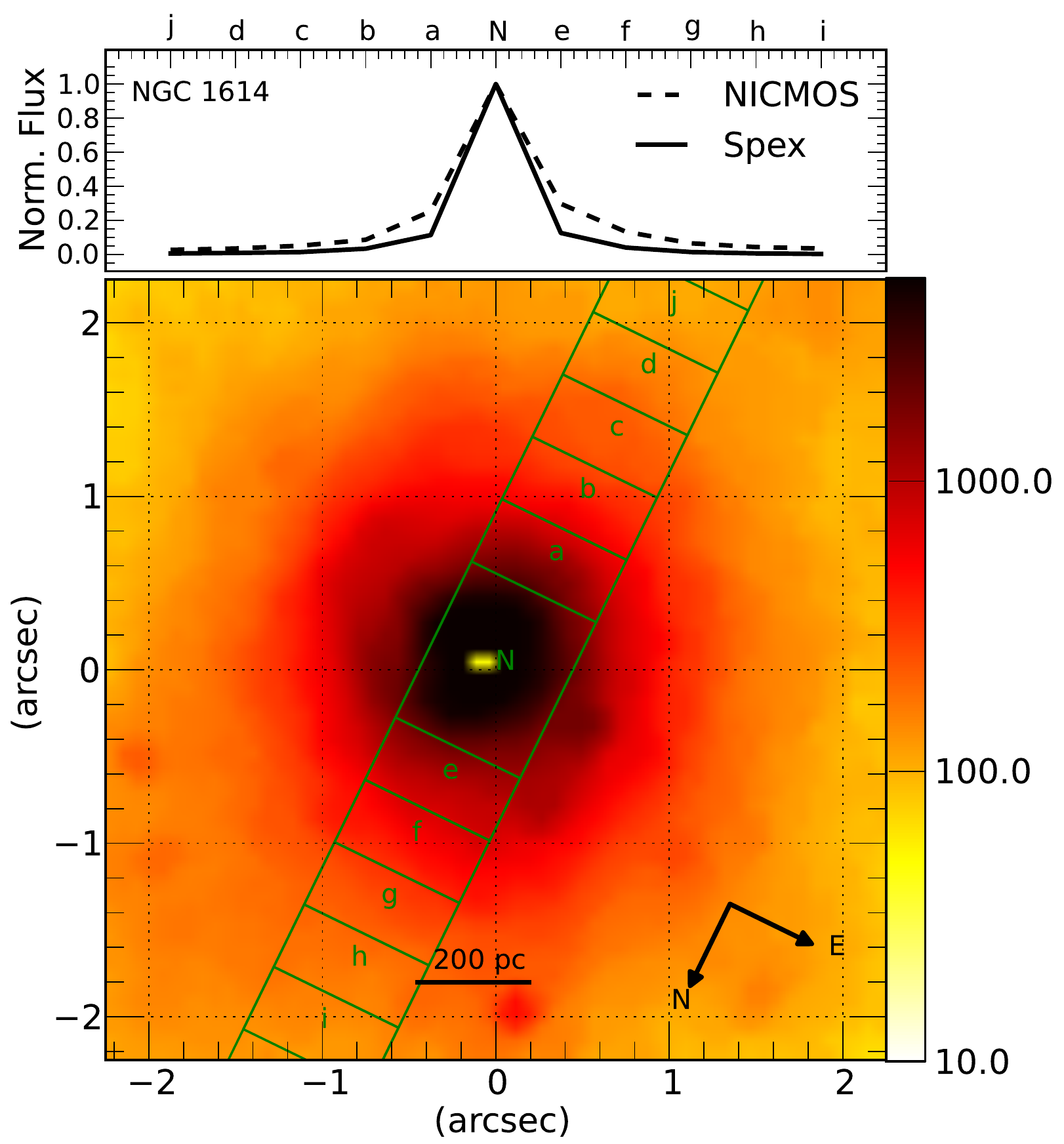}
\end{minipage} \hfill
\begin{minipage}[b]{0.49\linewidth}
\includegraphics[width=\linewidth]{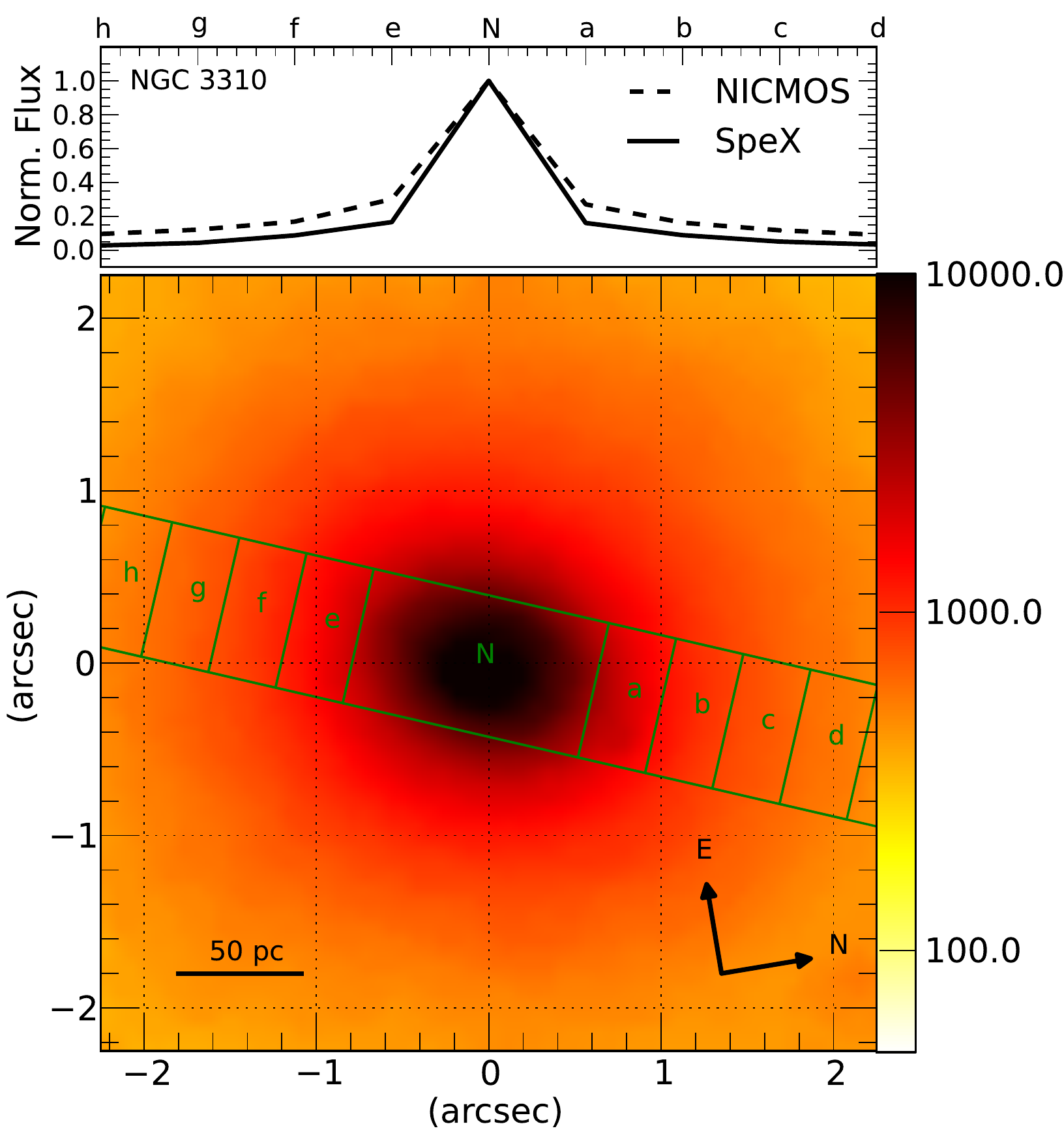}
\end{minipage} \hfill
\begin{minipage}[b]{0.49\linewidth}
\includegraphics[width=\linewidth]{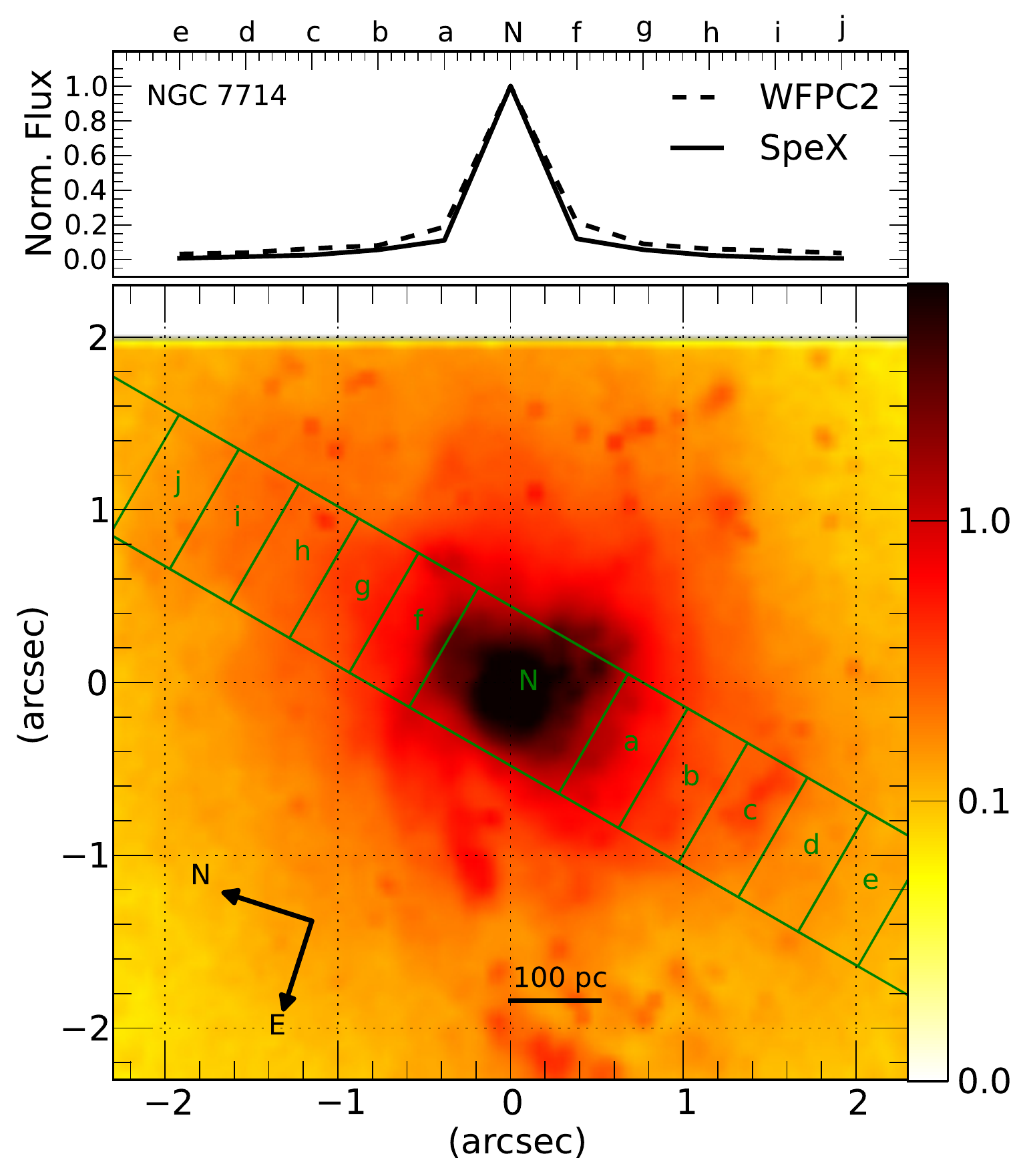}
\end{minipage} 
\caption{{\it Top:} NICMOS and SpeX profile (continuum profile at ${\lambda}_{cent}$=12230$\rm{\AA}$). {\it Bottom:} slit position overlapping the {\sc NICMOS} image of the galaxies. For NGC\,7714 we present 
WFPC2/F814W image.}
\label{slit}
\end{figure*}

Spectra of NGC\,1614, NGC\,34, NGC\,3310, NGC\,7714 were obtained at the NASA 3m Infrared
Telescope Facility (IRTF) in two observing runs. The first one
in April 21, 2002, and the second one
on the night of October 24, 2003. Table~\ref{obs:tab} shows
the log of observations for all galaxies. The
SpeX spectrograph \footnote{http://irtfweb.ifa.hawaii.edu/$\sim$spex/} \citep{ray03} was used in the short
cross-dispersed mode (SXD, 0.8-2.4 $\mu$m).  The detector
consists of a 1024x1024 ALADDIN~3 InSb array with a spatial scale
of 0.15$\arcsec$/ pixel. A 0.8$\arcsec$x 15$\arcsec$ slit oriented along the parallactic angle (see
Col.9 of Table~\ref{obs:tab}) was used, providing a spectral resolution, on average, of
320 km\,s$^{-1}$. This value was determined both from the arc lamp and 
the sky line spectra and was found to vary very little with wavelength along 
the observed spectra. The seeing, on average, 
was 0.8$\arcsec$. As SpeX does not provide the position of the slit centre, we derived them by 
matching the slit and the image profile using ${\chi}^2$ statistics.
The results are shown in Fig.~\ref{slit} where we present the slit position overlapping the 
NICMOS images of the galaxies. For NGC\,7714 we present the Wide Field and Planetary Camera 2 (WFPC2) F814W filter (${\lambda}=7940\rm{\AA} $) 
image, as there is no NICMOS image available for this galaxy. 

Observations were done nodding in an object-sky-sky-object  pattern,   
with the sky position usually several arcminutes from the galaxy nucleus,
free of extended emission or background stars.
Immediately after each galaxy a A0V telluric star, close in airmass to the former, 
was observed to remove telluric features and to perform the flux calibration, see \citet{rrp06} 
where a complete description of the reduction procedure is provided.
In summary, the spectral reduction, extraction and wavelength calibration
procedures were performed using {\sc spextool}\footnote{http://irtfweb.ifa.hawaii.edu/~cushing/Spextool.html}, the in-house
software developed and provided by the SpeX team for
the IRTF community \citep{cus04}.  In addition to the nuclear spectrum, 
a different number of off-nuclear apertures were extracted for each galaxy, 
depending on the size of the extended emission across the slit. 
Table~\ref{obs:tab} (Col.11) lists the diameter of the nuclear apertures. 
Off-nuclear spectra were extracted with 
a diameter of 0.4$\arcsec$ along the spatial direction at both sides of the
nuclear aperture until the signal drops to 1\% of the peak nuclear value. 

Telluric features removal and flux calibration were done using {\sc xtellcor} \citep{vacca03},
another software available by the SpeX team, which was designed specifically to 
perform telluric corrections on spectra obtained with Spex.
Thereafter, the different orders were 
merged into a single 1D spectrum from 0.8$\mu$m to 2.4$\mu$m using the 
{\sc xmergeorders} routine. The agreement between the overlapping region of two consecutive
orders was excellent and usually with errors of less than 1\%. Finally, the 
merged spectra were corrected for Galactic extinction using the \citet{ccm89} law
and the extinction maps of \citet{schlegel98}.
 
\begin{figure*}
\includegraphics[width=0.9\linewidth]{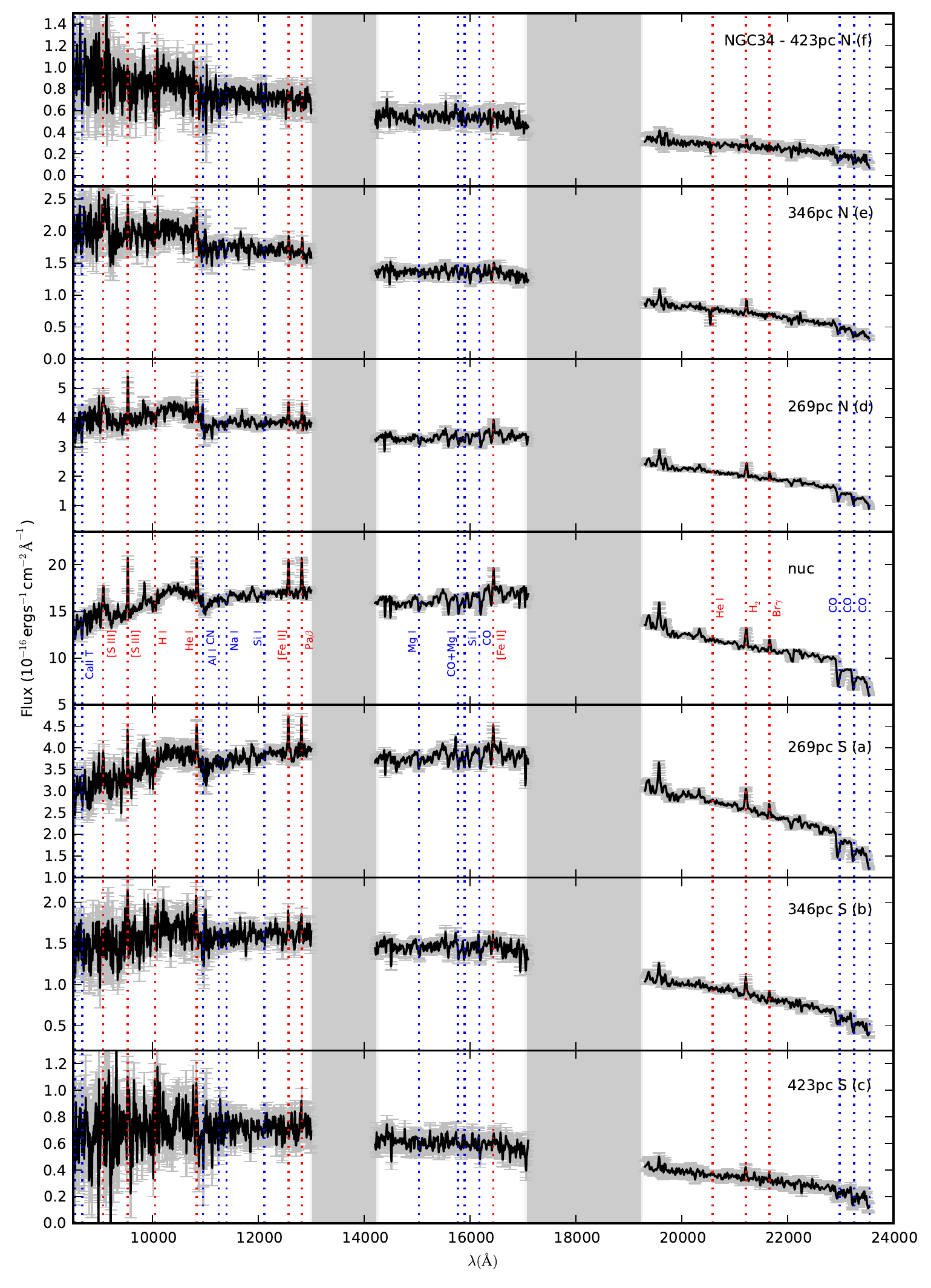}
\caption{NGC\,34 nuclear and extended region spectra. North (N) and south (S) are on the labels. The letters representing 
the apertures are used along the other figures (see Fig.~\ref{slit}). Error bars are in {\it gray}. Absorption ({\it blue}) and emission 
({\it red}) lines are marked. Telluric absorption regions are displayed in the shaded area {\it gray}.}
\label{especs34}
\end{figure*}

\begin{figure*}
\includegraphics[width=0.9\linewidth]{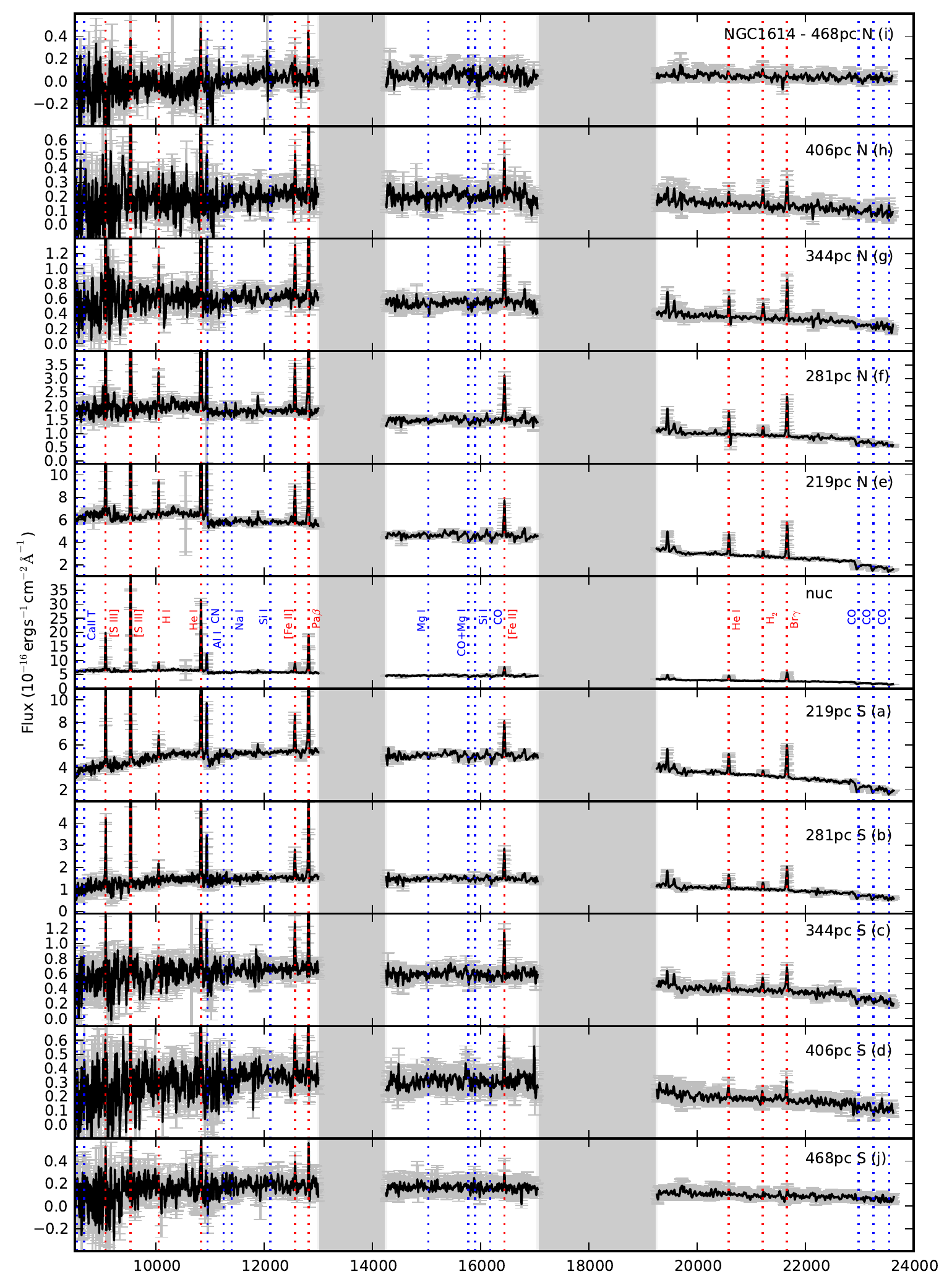}
\caption{Same as Fig.~\ref{especs34}, but for NGC~1614.}
\label{especs1614}
\end{figure*}

\begin{figure*}
\includegraphics[width=0.9\linewidth]{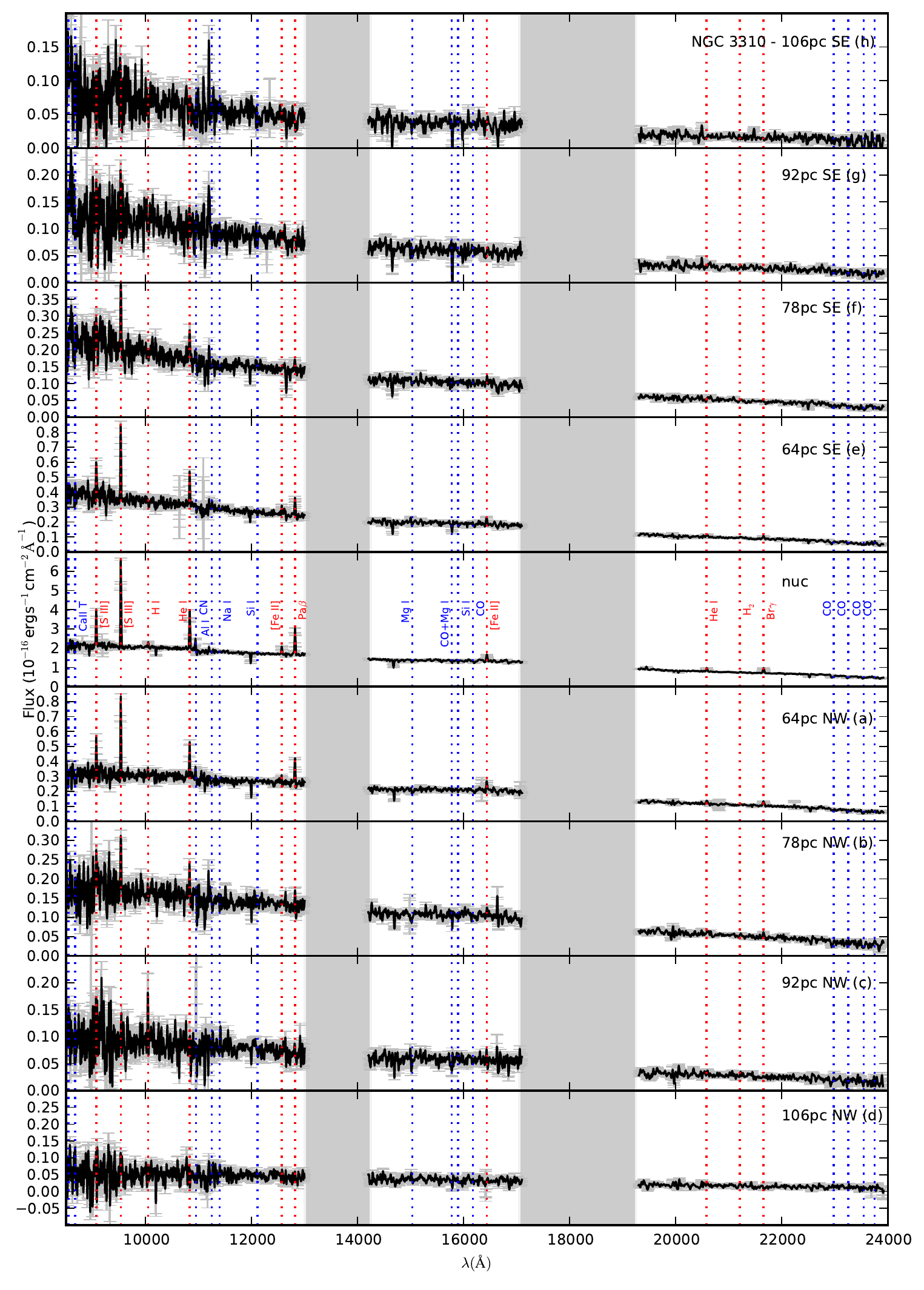}
\caption{Same as Fig.~\ref{especs34}, but for NGC~3310. Northwest (NW) and southeast (SE) are on the labels.}
\label{especs3310}
\end{figure*}

\begin{figure*}
\includegraphics[width=0.9\linewidth]{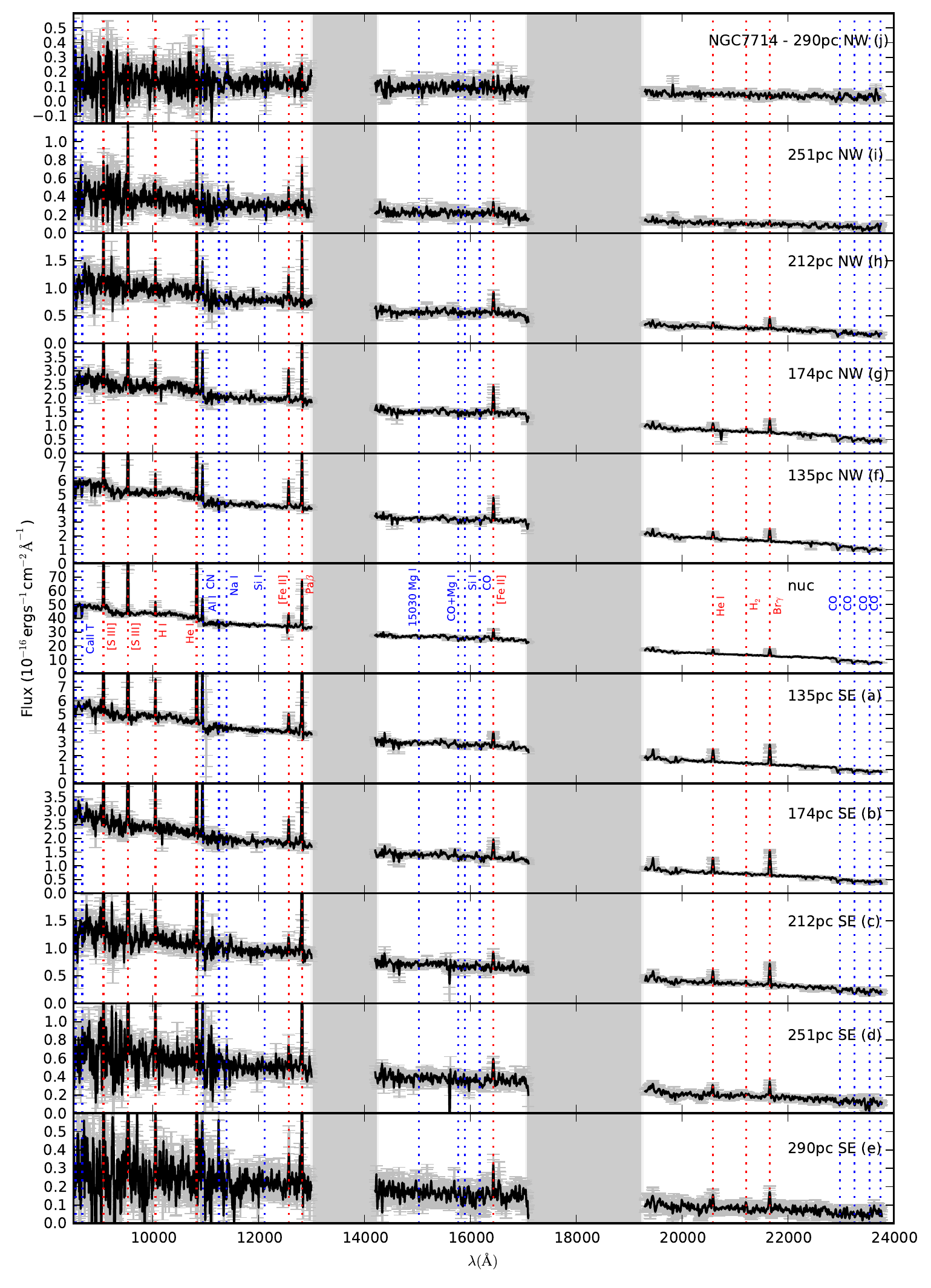}
\caption{Same as Fig.~\ref{especs34}, but for NGC~7714. Northwest (NW) and southeast (SE) are on the labels.}
\label{especs7714}
\end{figure*}

Figs~\ref{especs34} to \ref{especs7714} show the nuclear and off-nuclear spectra for each source, already
corrected by redshift. The most prominent absorption and emission lines are marked, as well as the telluric absorption regions.
The redshift, listed in Table~\ref{obs:tab} (Col.5), was determined from the position of the emission lines 
of [S\,{\sc iii}] 0.9531$\mu$m, He\,{\sc i} 1.083$\mu$m, Pa$\beta$ and Br$\gamma$ measured
in the nuclear spectrum of each galaxy. 
 
\begin{table*}
\centering
\caption{Log of observations}
\begin{tabular}{lccccccccccc}
\hline
Galaxy & Type & RA & DEC & z & Date of & T$_{\rm exp}$ & Air & PA$_{\rm obs}$ & {\it E(B-V)}$_G$ & Nuclear & Scale \\
 &      &  (h m s) & ($\deg$ ’ ”) & & observation & (s) & Mass & ($\deg$) & (mag) & Aperture ($\arcsec$) & (pc/$\arcsec$) \\
 (1)    &  (2)    & (3)       & (4)        &  (5)     &   (6)    &  (7) &  (8) & (9) & (10) & (11) & (12) \\
\hline
NGC 34   &SB/Sy2  & 00:11:06  &-12:06:26   & 0.019774 & 24/10/03 & 1680 & 1.19 & 4   & 0.027 & 1.0 & 383 \\
NGC 1614 & SB     & 04:33:59  & -08:34:46  & 0.016070 & 24/10/03 & 1800 & 1.14 & 0   & 0.154 & 1.0 & 324 \\
NGC 3310 & SB     & 10:38:45  & 00:33:06   & 0.003641 & 21/04/02 & 840  & 1.21 & 158 & 0.022 & 1.4 & 71 \\
NGC 7714 & H\,{\sc ii} & 23:36:14 & 02:09:18 & 0.009931 & 24/10/03 & 2400 & 1.05 & 348 & 0.052 & 1.0 & 192 \\
\hline
\end{tabular}
\begin{list}{Table Notes:}
\item (1) Galaxy name; (2) Galaxy classification; (3) Right ascension; (4) Declination; (5) Average redshift,
determined from the position of the emission lines of[S\,{\sc iii}] 0.9531$\mu$m, He\,{\sc i} 1.083$\mu$m, Pa$\beta$ and Br$\gamma$; 
(6) Date of observation; (7) Total exposure time for each galaxy; (8) Air mass; (9) Position angle of the slit; (10) Galactic
extinction from NED; (11) Aperture diameter for the nuclear extraction; Off-nuclear apertures were extracted 
with a diameter of 0.4$\arcsec$ until the extended emission drop to 1\% of the peak value; (12) Plate scale.
\end{list}
\label{obs:tab}
\end{table*}

\section{The stellar population synthesis method}\label{spsm}

There are two key ingredients when performing SP synthesis in galaxies: the base set (a set of simple stellar populations templates - SSPs) and the code.
The purpose is to disentangle the SP components from the spectral energy distribution of a galaxy into the different SPs contributions. For that, 
the code will mix the SSPs until it fits the galaxy's absorption spectra. Thus, an ideal base set should cover the range of spectral properties observed in the galaxy sample 
\citep{cid05} and should provide enough resolution in age and metallicity, in order to best fit the observed spectrum. 
In other words, a reliable base set would be a library of integrated spectra of star clusters 
\citep[i.e. they only depend on ages and metallicities of the stars and are free from any assumptions on stellar evolution and the 
initial mass function - ][]{bica86,riffel11a}. However, up to now there is no such library available in the literature for the NIR 
spectral region. In this way, the use of a base set composed of theoretical SSPs, covering this spectral region, is becoming a common approach 
\citep[e.g.][]{riffel09,thaisa12,martins10,martins13}. 

\begin{figure*}
\includegraphics[width=\linewidth]{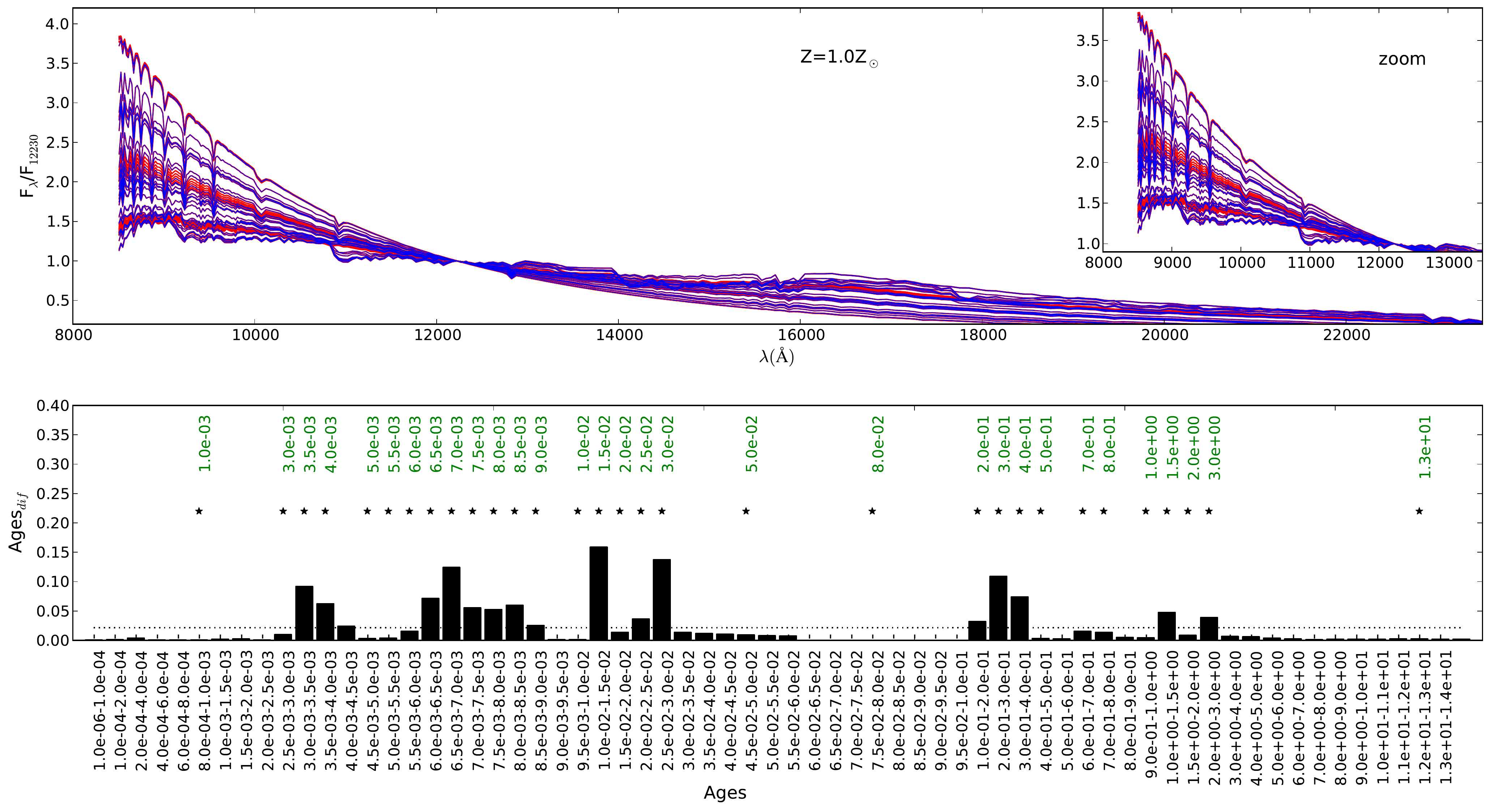}
\caption{All the 67 M05 SSPs for Z=Z$_{\odot}$ are plotted ({\it red}) in the upper panel together with the 31 SSPs selected ({\it blue}) to compose the base set for this metallicity. 
In the bottom panel the quadratic difference between two consecutive SSPs with consecutive ages. The representative templates chosen to compose the base of elements 
are marked with a star and the correspondent ages are plotted in {\it red} in the top panel. A zoom of the blue end (where the main 
differences are observed) is also shown.}
\label{base0.5}
\end{figure*}

Since the NIR carries fingerprints from evolved stars \citep[e.g.][]{riffel07,ramos09,martins13} and these are crucial to model the 
absorption line spectra of the galaxies, we used the M05 EPS models. They include empirical spectra of C- and O-rich stars 
\citep{lw00} and thus, are able to predict these features. These models (gray lines in Fig.~\ref{base0.5}) span over an age range from 1\,Myr to 15\,Gyr 
according to a grid of 67 models with four different metallicities\footnote{It is worth calling attention to the reader that Maraston 
do provide models with Z=0.005Z$_{\odot}$ and Z=3.5Z$_{\odot}$, but only for ages older than 1\,Gyr, therefore we left them out of the base set.}. Since, 
small differences in the stellar population are washed away in real data \citep[i.e. they are diluted by the noise in real data][]{cid04} 
we decided only to include in the base the representative SSPs (i.e. where significant differences between them are observed) to avoid 
redundant informations and degeneracies in the base set. In order to choose the elements to compose the base we compute the square root 
of the quadratic difference between two consecutive SSPs with consecutive ages (t) normalized by the number of pixels (N), i.e. solving the equation: 

\begin{equation}
Ages_{dif} = \frac{1}{N}\sum_{t_0}^{t_f}\sqrt{(F_{\lambda}(t_i) - F_{\lambda}(t_{i+1}))^2}
\end{equation}

An example of the results of this procedure, for the solar metallicity, is shown in Fig.~\ref{base0.5}. The representative SSPs 
(which constitute the base set) are marked with a star and are overplotted in red on Fig.~\ref{base0.5} top. Note that this procedure 
was done only in the wavelength interval between 8000$\rm{\AA}$-24500$\rm{\AA}$ and for the four metallicities with all ages available. The final base 
set results in 31 ages, t=0.0010, 0.0030, 0.0035, 0.0040, 0.0050, 0.0055, 0.0060, 0.0065, 0.0070, 0.0075, 0.0080, 0.0085, 0.0090, 0.010, 0.015, 0.020, 0.025, 0.030, 
0.050, 0.080, 0.2, 0.3, 0.4, 0.5, 0.7, 0.8, 1,1.5, 2, 3, 13 Gyr, and 4 metallicities, Z= 0.02, 0.5, 1 and 2 $Z_{\odot}$. Thus, the final base set is composed of 124 SSPs. 

\begin{figure*}
\includegraphics[width=\linewidth]{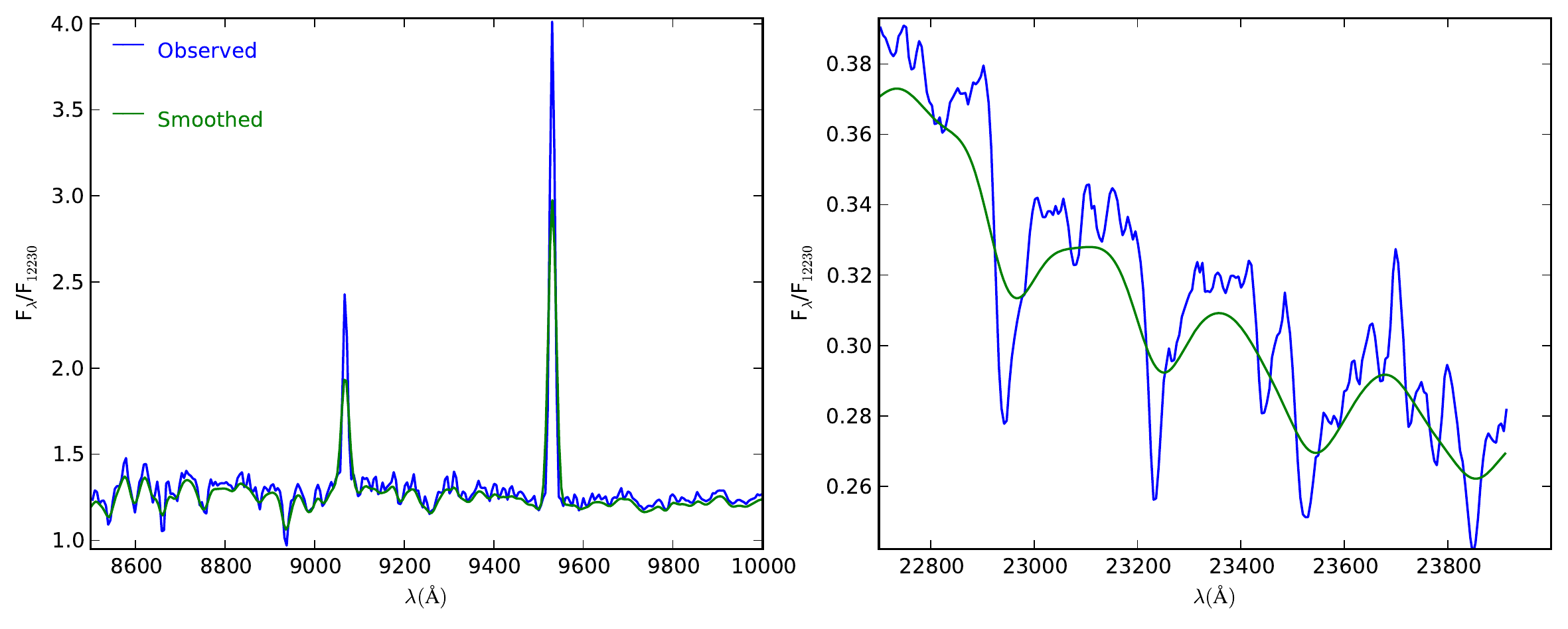}
\caption{Comparison between the observed nuclear spectrum of NGC\,3310 ({\it blue}) and the result of the gaussian smoothing ({\it green}). We split the spectrum
in two parts, the {\it left panel} presents the effect of the gaussian smoothing on the emission lines and the {\it right panel} shows the effect on the CO absorption band.}
\label{comp_smt}
\end{figure*}

It is important to call the attention to the fact that the spectral resolution of the M05 models in the NIR is significantly lower 
(R $\leq$ 250) than that of the observed data \citep[R$\sim$ 750, ][]{ray03} and varies with wavelength. For this reason, the observations 
were degraded to the model’s resolution, by convolving them with a gaussian. One example of a final smoothed spectrum 
is shown in Fig.~\ref{comp_smt}.

Once the base set is defined, the other fundamental ingredient in SP fitting is the code, which will mix the individual components 
of the base set to get the best fit to the observed spectrum. Here we used the {\sc starlight} code 
\citep{cid04,cid05,cid09,cid10,cid11,mateus06,asari07,asari09} which mixes computational techniques originally developed 
for empirical population synthesis with ingredients from EPS models. Briefly, the code fits the observed spectrum {\it $O_{\lambda}$} 
with a combination in different proportions of {\it $N_{\star}$} SSPs in the base set, {\it $b_{j, \lambda}$} taken from the EPS models. 
Another important aspect is that the code fits the entire spectrum, from 0.8 to 2.4 $\mu$m, excluding emission lines and spurious features 
(e.g. cosmic rays, noise and telluric regions), which are masked out or clipped (see below for details). 
 
The extinction law used in this work was the Calzetti law \citep{calzetti00} implemented by Hyperz \citep{bol00}, a public photometric redshift code
which computed the Calzetti law for $\lambda >$  2.2$\mu$m.

Velocity dispersion is a free parameter for {\sc starlight}, which broadens the SSPs in order to better fit the absorption lines in the observed
spectra. However this step is not relevant in our case, because of the low resolution models.

Basically, {\sc starlight} solves the following equation for a model spectrum {\it $M_{\lambda}$} \citep{cid05}:

\begin{equation}
{\it M_{\lambda}= M_{\lambda 0} \left[ \sum_{j=1}^{N_{\star}} x_j b_{j, \lambda} r_{\lambda} \right] \otimes G(v_{\star}, {\sigma}_{\star})}
\end{equation}

\noindent where ${\it M_{\lambda 0}}$ is the synthetic flux at the normalization wavelength ({\it $\lambda_0=12240\rm{\AA}$}); ${\it x_j}$ is 
the {\it j}th population vector component of the base set; ${\it b_{j, \lambda} r_{\lambda}}$ is the reddened 
spectrum of the {\it j}th SSP normalized at ${\lambda}_0$ in which
${\it r_{\lambda}=10^{-0.4(A_{\lambda}-A_{\lambda 0})}}$ is the extinction term; ${\it \otimes}$ denotes 
the convolution operator and ${\it G(v_{\star}, {\sigma}_{\star})}$ is the gaussian distribution used to model 
the line-of-sight stellar motions, centred at velocity ${\it v_{\star}}$ with dispersion ${\it {\sigma}_{\star}}$.
We choose as normalization wavelength {\it $\lambda_0=12240\rm{\AA}$}, since this region is free from emission and absorption lines. {\sc starlight} normalizes all the base spectra by {\it F$_{\lambda_0}$}
and the observed spectrum is normalized by the mean flux measured in a window (we used 80$\rm{\AA}$), a region centred at 
{\it $\lambda_0$} defined in order to avoid the effect of a bad pixel (noise spike, cosmic ray, etc.) at {\it $\lambda_0$} in the 
observed spectrum. Finally, the best fit is achieved by {\sc starlight} as the code searches for the minimum of the equation:

\begin{equation}
\label{chi2}
{\it {\chi^2 = \sum_{\lambda}^{} {\left[ \left( {\it O_{\lambda}-M_{\lambda}}\right)w_{\lambda} \right]}^2 }}
\end{equation}

The {\sc starlight} output parameters ${\chi}^2$ and Adev can be used to measure the robustness of the SP fit. Adev is the percent 
mean deviation $|O_{\lambda} - M_{\lambda}|/O_{\lambda}$, where $O_{\lambda}$ is the observed spectrum and $M_{\lambda}$ is the 
fitted model. The fits are carried out with a mixture of simulated annealing \citep{kirk83} plus Metropolis scheme, which gradually 
focuses on the most likely region in parameter space, avoiding (through the logic of the cooling schedule) trapping on to local minimum. 

The emission lines and spurious features (noise, telluric regions, cosmic rays) are masked out by using ${\it w_{\lambda}=0}$ in 
the regions where they are located. In our case, the emission lines masked were [S{\sc iii}] 9069$\rm{\AA}$, [S{\sc iii}] 9531$\rm{\AA}$, 
He{\sc i} 10830$\rm{\AA}$, Pa$\gamma$ 10938$\rm{\AA}$, [Fe {\sc ii}] 12570$\rm{\AA}$, Pa$\beta$ 12810$\rm{\AA}$, [Fe {\sc ii}] 16444$\rm{\AA}$, 
Pa$\alpha$ 19570$\rm{\AA}$, Br$\gamma$ 21650$\rm{\AA}$ and H$_{\rm 2}$ 21213$\rm{\AA}$. Spurious data were also masked out. Examples of masked regions are indicated in 
Figs~\ref{hp34} to \ref{hp7714}. 

In summary, after excluding the masked points, the minimization process consists of the following stages: (i) make a broad sweep of 
the parameter space, starting from a high `temperature' (in statistical mechanics terms) to allow the system to explore all sorts 
of configuration and then implementing a cooling schedule to gradually reduce the `temperature', therefore letting the system to 
settle into the lowest energy states, (ii) exclude pixels which could not be fitted by the first stage, once they deviate strongly
from the best M$_{\lambda}$ found in this first stage, (iii) detailed fit with the full base and (iv) the whole fit is fine-tuned 
repeating the full loop excluding all irrelevant components ($x_j$=0) to the fit. A well detailed discussion of the {\sc starlight} 
procedures can be found in \citet{cid04,cid05} and in {\sc starlight} manual, available at http://astro.ufsc.br/starlight/.

\subsection{Uncertainties on the Fits}\label{error}

The use of statistics in order to interpret spectral synthesis results is widespread \citep[e.g.][]{pan07}, once averaging results tends 
to reduce the uncertainties \citep{cid13}. Another good reason to use the statistical interpretation is to produce an estimation of the 
uncertainties, which {\sc starlight} does not provide as an standard output. The most straightforward way to achieve this is to 
perturb the input spectra according to some realistic prescription of the errors involved \citep{cid13}. In order to estimate these 
uncertainties, we simulate 100 spectra for each aperture of each galaxy in our sample. The simulated flux for each wavelength ($\lambda_i$) 
is computed assuming a Gaussian distribution of the uncertainties. Therefore, the mean flux in each $\lambda_i$ corresponds 
to the measured flux value and the standard deviation is given by the errors provided by {\sc spextool}. Thus, we simulate the 
flux at $\lambda_i$ according to its Normal probability.

{\sc starlight} provides a single best fit set of parameters among typically many millions trials during its likelihood guided sampling 
of the parameter space. These single solutions, however, are not necessarily mathematically unique. In this scenario, we performed SP synthesis 
in all the simulated spectra in order to obtain an average result together with the standard deviation associated to 
each aperture of each galaxy. As a result, we have an estimative of the uncertainties associated with the SP fitting (see Figs~\ref{hp34} to \ref{hp7714}).

\section{Results and Discussion}\label{resdisc}

\begin{figure*}
\includegraphics[width=0.9\linewidth]{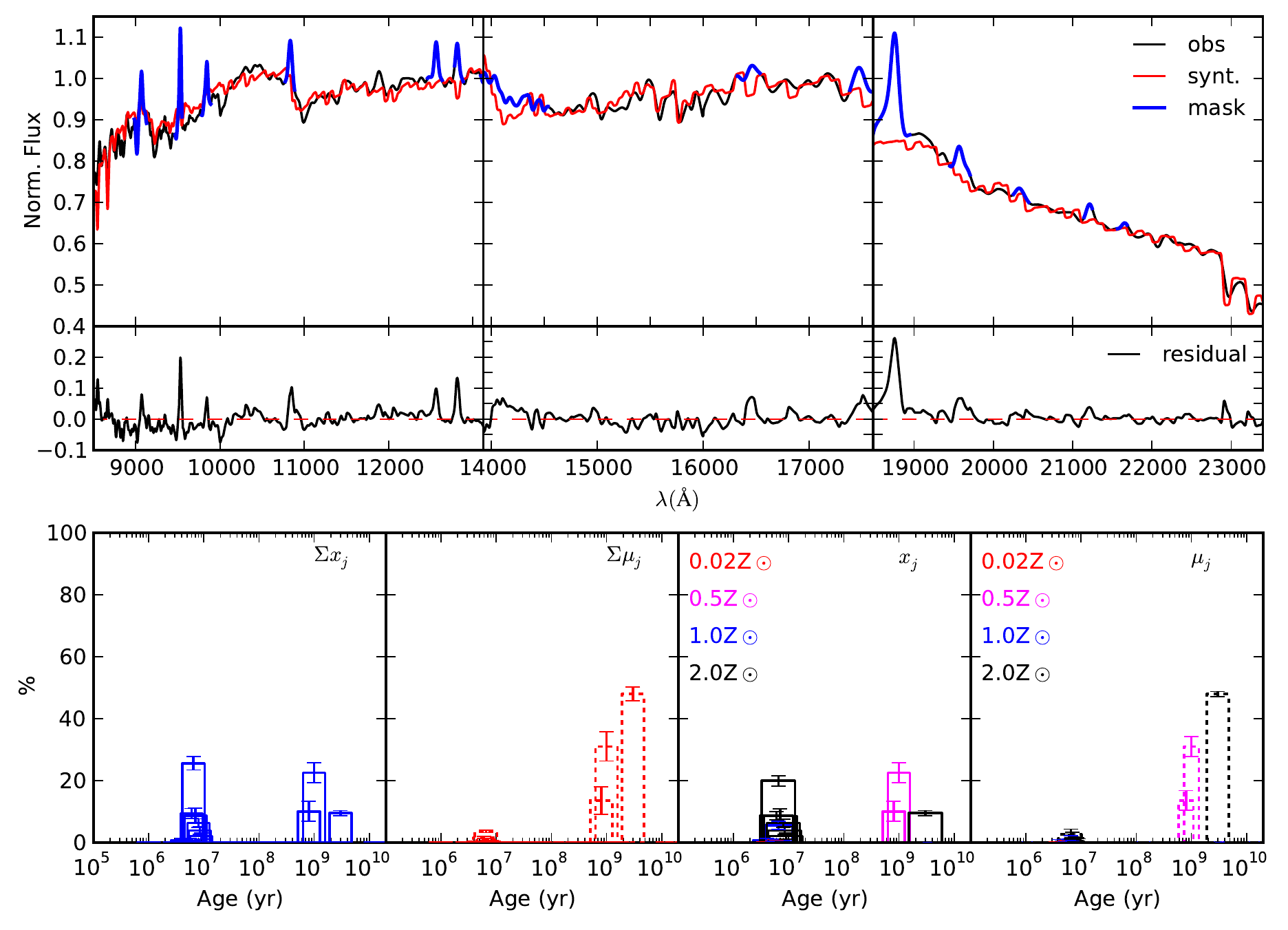}
\caption{Results for NGC\,34. {\it Upper panels:} we present the synthesis result ({\it red}) and the masked points ({\it blue}) overlapping 
the observed spectra ({\it black}), below the residual spectrum is shown (the dotted line in {\it red} marks the zero point in flux). 
{\it Bottom panels:} we display four histograms, the two on the left 
present the flux-weighted ($x_j$) and mass-weighted (${\mu}_j$) SPs vectors contributions sorted only by age (metallicities summed) and the two on the right show the 
flux-weighted ($x_j$) and mass-weighted (${\mu}_j$) SPs vectors contributions sorted by age and metallicity. Telluric regions were omitted.}
\label{hp34}
\end{figure*}

\begin{figure*}
\includegraphics[width=0.9\linewidth]{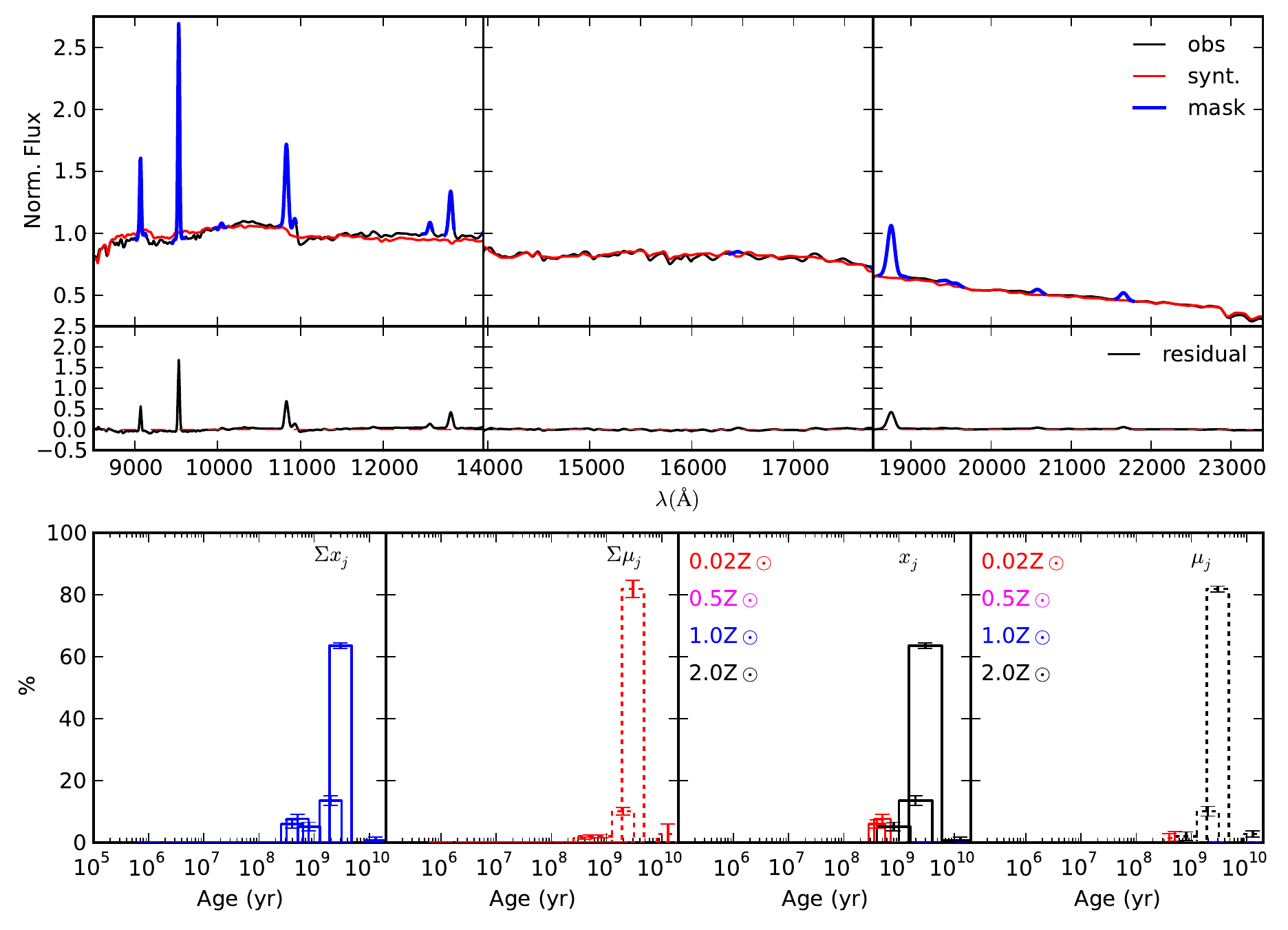}
\caption{The same as Fig.~\ref{hp34} but for NGC\,1614 }
\label{hp1614}
\end{figure*}

\begin{figure*}
\includegraphics[width=0.9\linewidth]{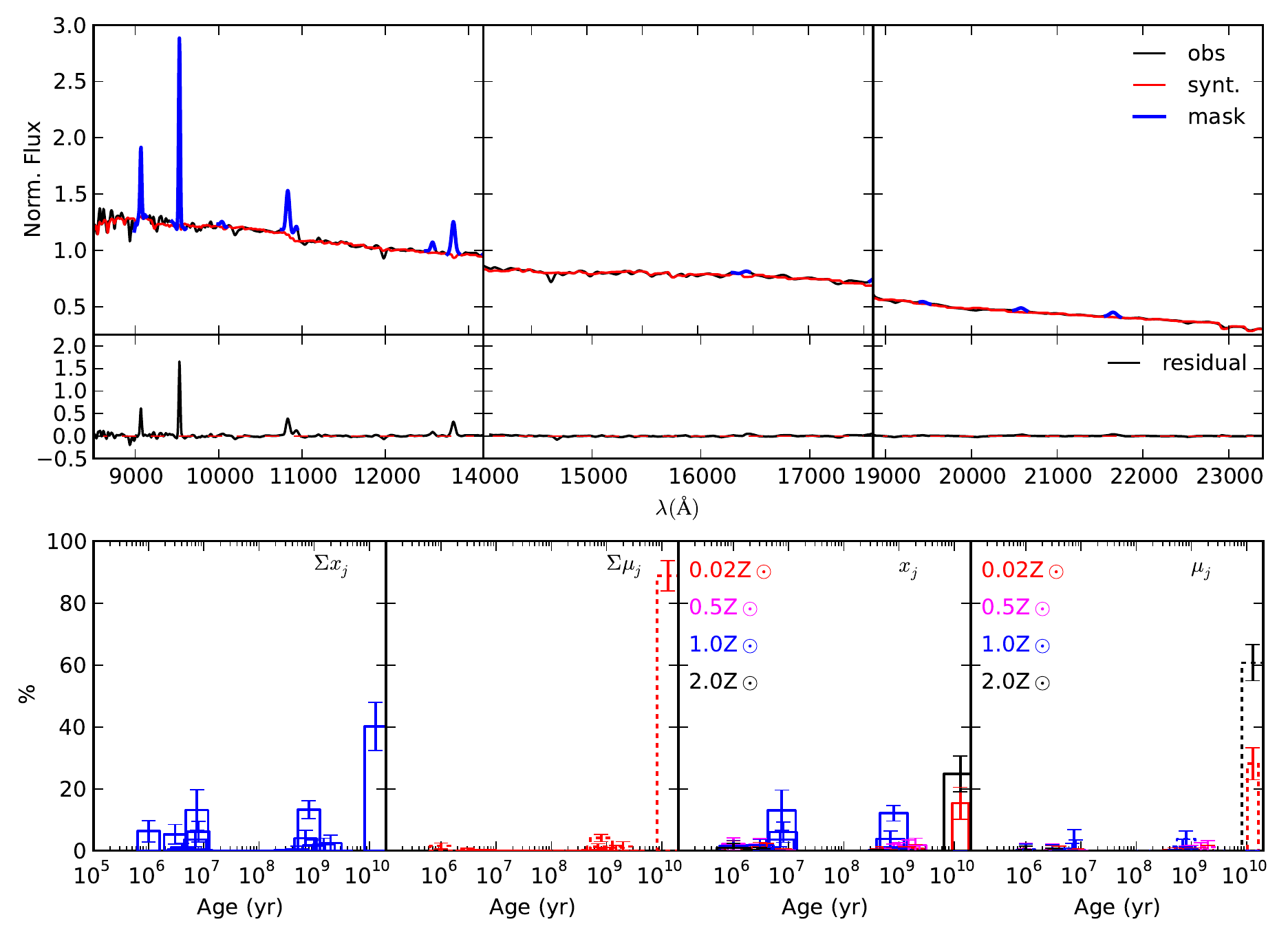}
\caption{The same as Fig.~\ref{hp34} but for NGC\,3310 }
\label{hp3310}
\end{figure*}

\begin{figure*}
\includegraphics[width=0.9\linewidth]{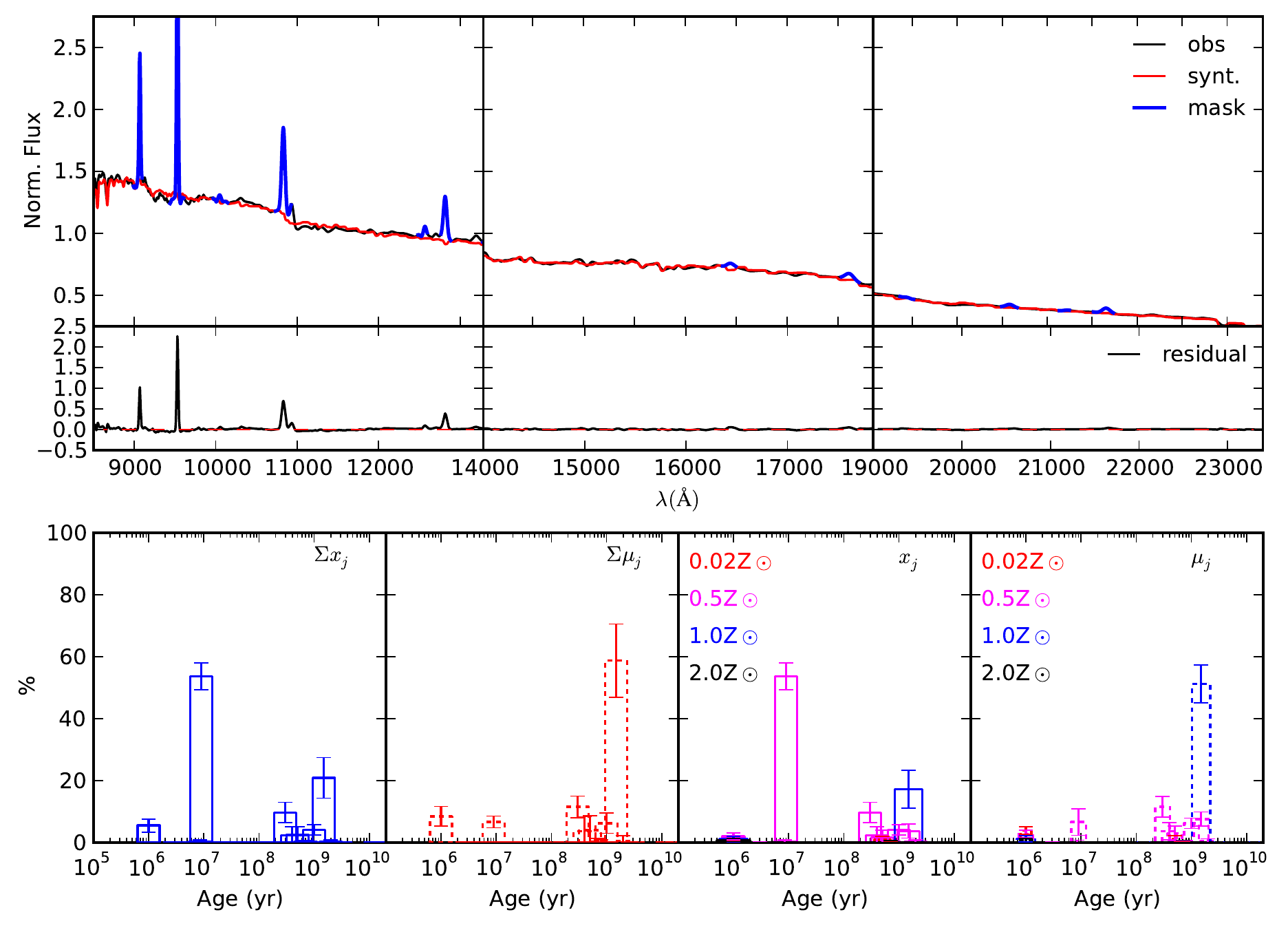}
\caption{The same as Fig.~\ref{hp34} but for NGC\,7714 }
\label{hp7714}
\end{figure*}

By making a systematic study over the SP variance along the galaxy, we built a scenario for the star formation in the four SBs studied here. 
An example of the individual results for the nuclear extraction of each galaxy is 
shown in Figs~\ref{hp34} to \ref{hp7714}. These plots include: (i) in the top panel are the synthesis 
result ({\it red}) and the masked points ({\it blue}) overlapping the observed spectra ({\it black}); (ii) the residual spectrum (the dotted line in {\it red} marks the 
zero point in flux) and (iii) the bottom panel shows four histograms, the two on the left present the flux-weighted 
($x_j$) and mass-weighted (${\mu}_j$) SP vectors contributions sorted only by age (metallicities summed) and the two on the right show $x_j$ and 
${\mu}_j$ sorted by age and metallicity.  
In these figures we can analyze the results individually for each aperture of each galaxy, controlling the quality of the fits and improving the mask files when necessary.

On the other hand, grouping the population vector in age bins should thus provide a coarser but more powerful description of the SFH of the galaxies
\citep[][, among others.]{cid01, cid03, riffel10, riffel11b}. Actually, \citet{cid01} show that measurement errors as well as the use of reduced sets of observables
are responsible for spreading a strong contribution in one component preferentially among base elements of same age. 
In this sense, we used their definition of binned population vectors as follows: \textit{young}: $x_y$ ($t \leq 50\times10^6$ yr), \textit{intermediate-age}: 
$x_i$ ($50\times10^6 < t \leq 2\times10^9$ yr) and \textit{old}: $x_o$ ($t > 2\times10^9$ yr).

In addition to this, in order to describe the metallicity behavior of the SP mixture along the galaxy, we have used the flux- and mass-weighted mean 
metallicity defined by \citet{cid05} as: 

\begin{equation}\langle Z_{\star} \rangle_{F} = \displaystyle \sum^{N_{\star}}_{j=1} x_j Z_j,\end{equation} for the flux-weighted mean metallicity and,
\begin{equation}\langle Z_{\star} \rangle_{M} = \displaystyle \sum^{N_{\star}}_{j=1} m_j Z_j \end{equation} for the mass-weighted mean metallicity.
\noindent
Both definitions are bounded by the $\frac{1}{50}Z_{\odot}$-2$Z_{\odot}$ range. 

\begin{figure*}
\includegraphics[width=0.9\linewidth]{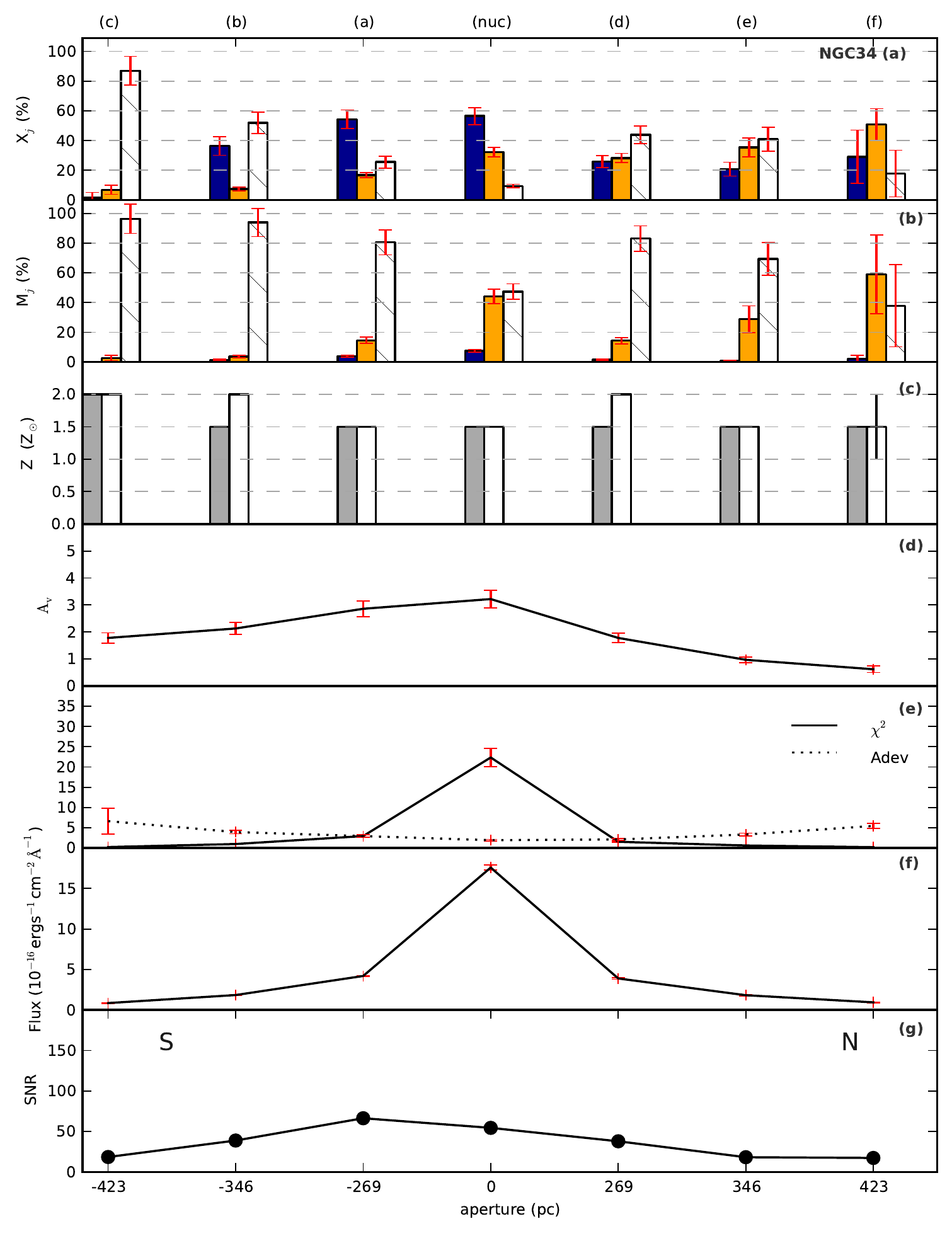}
\caption{Global analysis for NGC\,34. Panels {\it a} and {\it b} present the average contribution in flux and mass, respectively, of the population vectors along the apertures. 
Blue, yellow and red represent young ($x_y$), intermediate ($x_i$) and old ($x_o$) SP contributions respectively; 
panel {\it c} shows the average metallicity weighted by flux ($Z_F$ - {\it filled}) and by mass ($Z_M$ - {\it empty});
the extinction ($A\rm_{v}$) is plotted along the apertures in panel {\it d}; the average ${\chi}^2$ and Adev are plotted in panel {\it e};  
the continuum profile at ${\lambda}_{cent}=12230\rm{\AA}$ in panel {\it f} and for last, the signal-to-noise ratio (SNR) along the galaxy in panel {\it g}. 
The letters on the top of the figure correspond to those of Fig.~\ref{slit}. North (N) and south (S) direction are indicated in the bottom panel.}
\label{hist34}
\end{figure*}

\begin{figure*}
\includegraphics[width=0.9\linewidth]{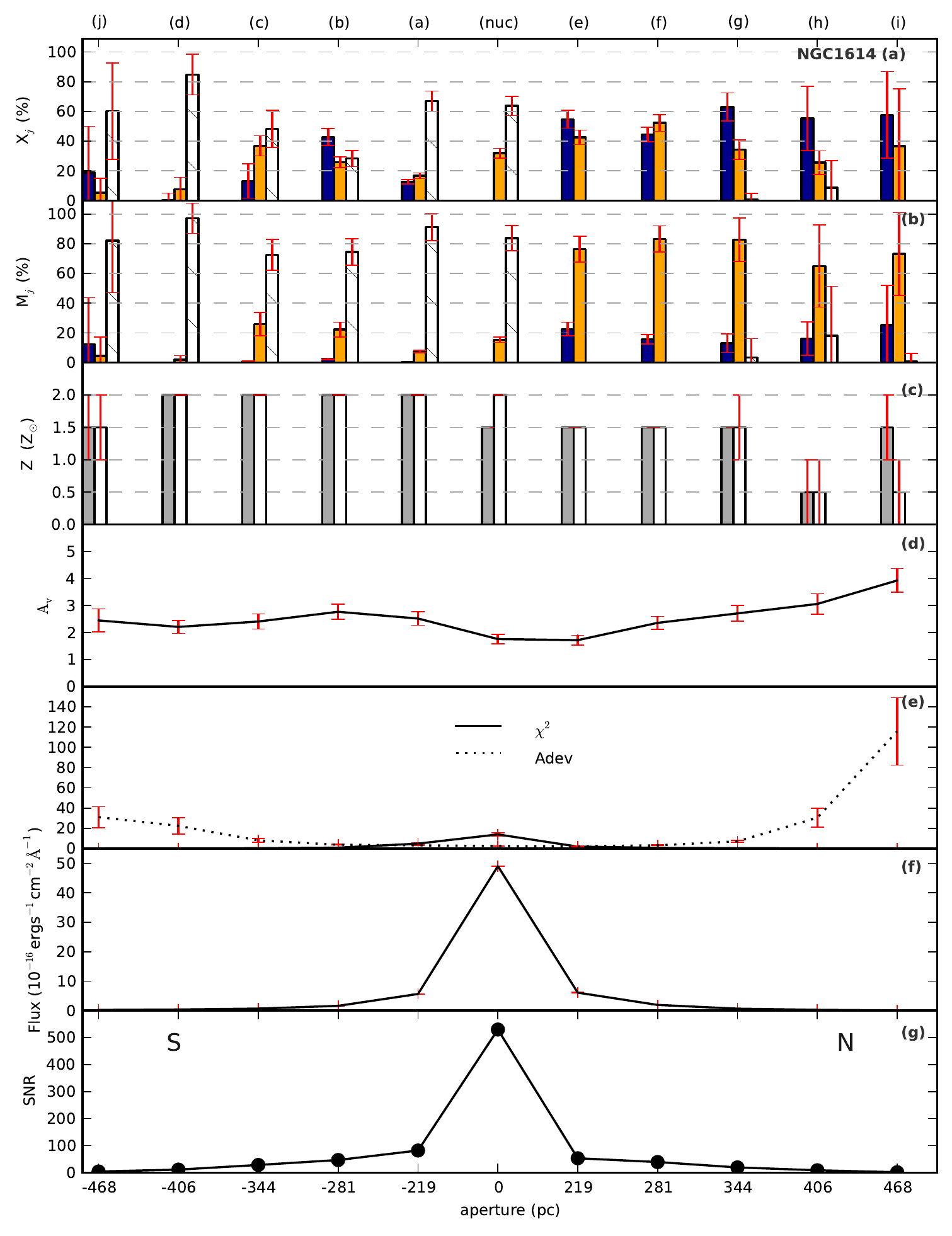}
\caption{The same as Fig.~\ref{hist34}, but for NGC\,1614. North (N) and south (S) direction are indicated in the bottom panel.}
\label{hist1614}
\end{figure*}

\begin{figure*}
\includegraphics[width=0.9\linewidth]{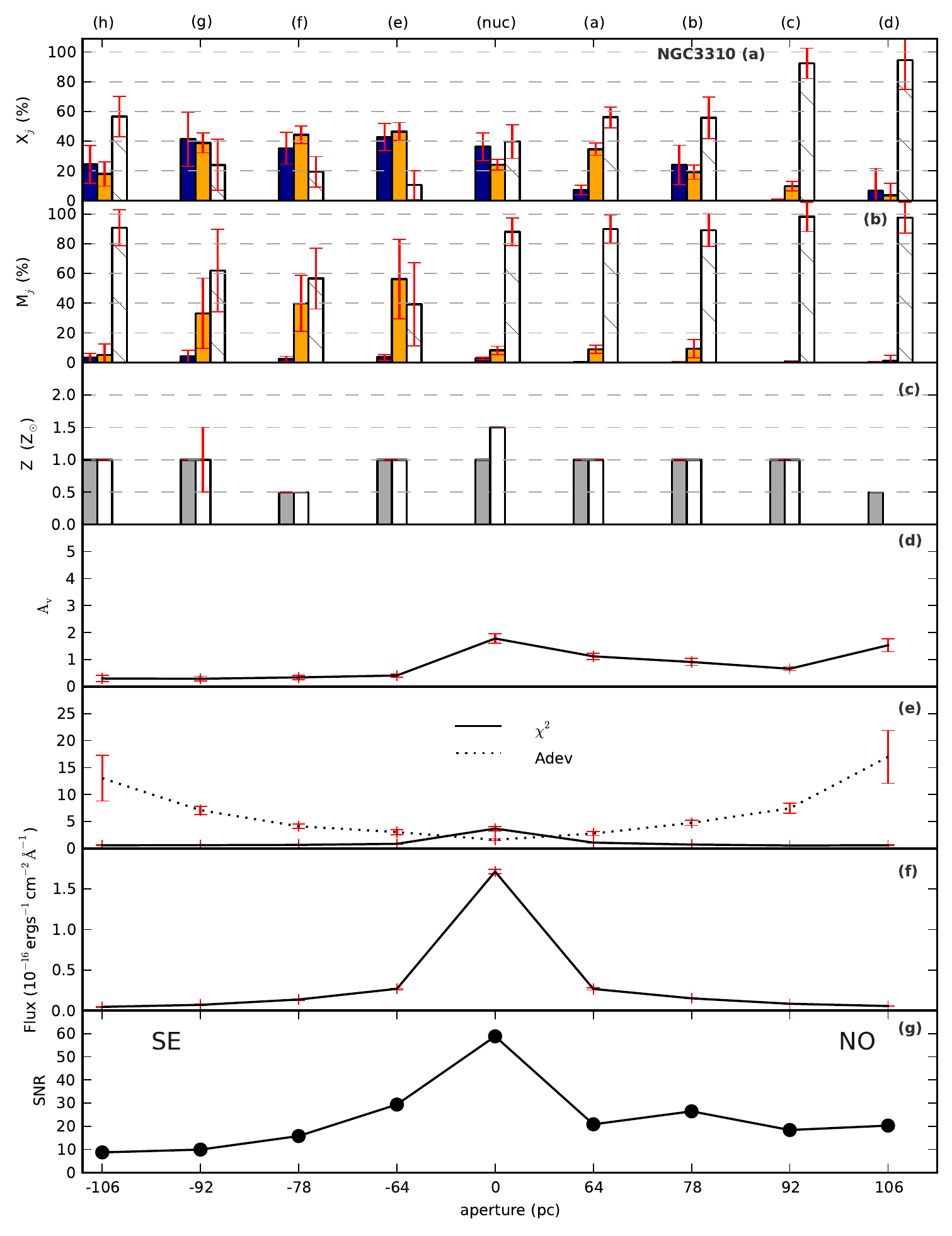}
\caption{The same as Fig.~\ref{hist34}, but for NGC\,3310. Northwest (NW) and southeast (SE) direction are indicated in the bottom panel.}
\label{hist3310}
\end{figure*}

\begin{figure*}
\includegraphics[width=0.9\linewidth]{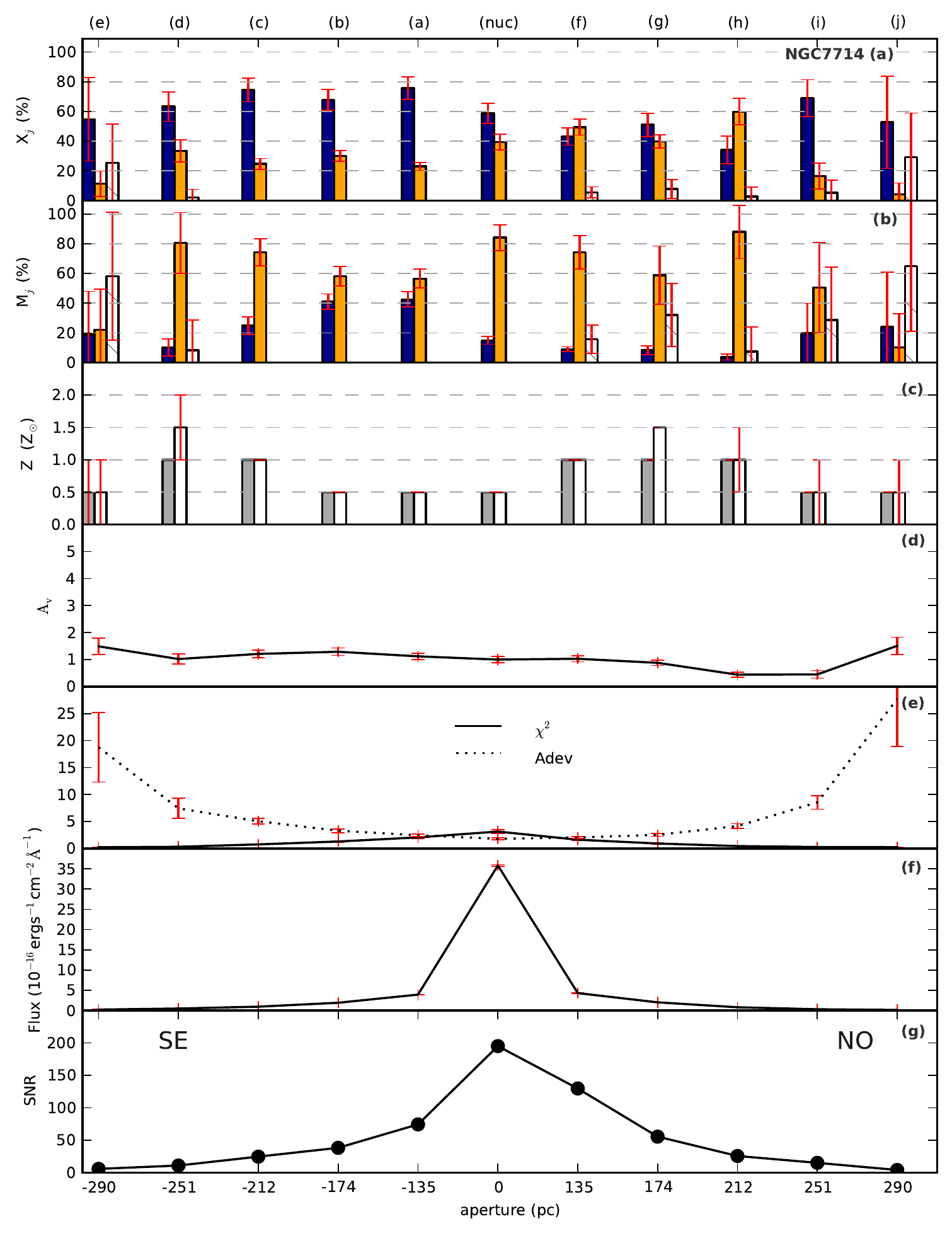}
\caption{The same as Fig.~\ref{hist34}, but for NGC\,7714. Northwest (NW) and southeast (SE) direction are indicated in the bottom panel.}
\label{hist7714}
\end{figure*}

Therefore, to best analyze our results we create a series of histograms for each galaxy shown in Figs~\ref{hist34} to \ref{hist7714}. 
Each figure includes 7 plots: (i) panels {\it a} and {\it b} present a global analysis of each source by plotting the average contribution 
in flux and mass, respectively, of the binned vectors along the apertures; (ii) panel {\it c} shows the mean metallicity weighted by 
flux ({\sc $Z_F$}) and by mass ({\sc $Z_M$}); (iii) the extinction ($A\rm_{v}$) is plotted along the apertures in panel 
{\it d}; (iv) the average ${\chi}^2$ and Adev are plotted in panel {\it e}; (v) the continuum profile at ${\lambda}_{cent}=12230\rm{\AA}$ is 
shown in panel {\it f}, (vi) The signal-to-noise ratio (SNR) in panel {\it g}, which was estimated as being the ratio between the mean values
of the F$_{\lambda}$ points (in a window of 60$\rm{\AA}$) and their standard deviation. These histograms will enable us to construct the star 
formation scenario for each galaxy, revealing the predominant ages of the population as well as indicating the possible presence of nuclear structures.

It is important to mention that the uncertainties increase with decreasing SNR (higher error bars in the outer apertures - see Figs~\ref{hist34} to \ref{hist7714}). 
Nevertheless, we decided to include these results (for SNR $\leq$10) in the analysis, as they follow the same tendency of the SP distribution along the galaxy. 
The average results from the SP synthesis are in Table~\ref{sr}.

\begin{table*}
\centering
\scriptsize
\caption{Average synthesis results.}
\label{sr}
\begin{tabular}{lllllllllllll}
\noalign{\smallskip}
\hline
\hline
\noalign{\smallskip}
 Galaxy  & Aperture\,(pc) & {\sc X$_y$}\,(\%)  & {\sc X$_i$}\,(\%)  & {\sc X$_o$}\,(\%)  & {\sc M$_y$}\,(\%)  & {\sc M$_i$}\,(\%)  & {\sc M$_o$}\,(\%)  & {\sc Z$_L$}\,({\sc Z$_{\odot}$}) & {\sc 
Z$_M$}\,({\sc Z$_{\odot}$}) & $A\rm_{v}$\,(mag) & Adev\,(\%) & ${\chi}^2$\\
         &   &    (1)  &  (2)    &  (3)   &(4)    & (5)  & (6)   &  (7)	&  (8)  & (9) & (10)  & (11)  \\
\noalign{\smallskip}
\hline
\hline
\noalign{\smallskip}  
NGC\,34   & (c)423\,S &  2\pp5     &  7\pp3     &   87\pp11  &  ...   &  3\pp2   & 96\pp10 &  2.0  &   2.0	  &  1.79\pp0.19   &  6.3\pp1.6   &  0.2       \\  
          & (b)346\,S & 36\pp6     &  7\pp1     &   52\pp7   &  1     &  4\pp1   & 94\pp9  &  1.5  &   2.0	  &  2.13\pp0.22   &  3.9\pp0.4   &  1.0\pp0.1 \\
          & (a)269\,S & 54\pp6     &  17\pp2    &   26\pp4   &  4\pp1 &  15\pp2  & 81\pp8  &  1.5  &   1.5	  &  2.86\pp0.29   &  2.9\pp0.3   &  2.9\pp0.3 \\ 
          &    Nuc    & 49\pp5     &  31\pp3    &   18\pp2   &  5\pp1 &  30\pp3  & 64\pp7  &  1.5  &   1.5	  &  3.23\pp0.32   &  2.2\pp0.2   & 30.4\pp3.1 \\
          & (d)269\,N & 26\pp4     &  28\pp3    &   44\pp6   &  2     &  14\pp2  & 83\pp9  &  1.5  &   2.0	  &  1.78\pp0.18   &  2.1\pp0.2   &  1.5\pp0.2 \\    
          & (e)346\,N & 21\pp5     &  35\pp6    &   41\pp8   &  1     &  29\pp9  & 69\pp11 &  1.5  &   1.5	  &  0.97\pp0.10   &  3.3\pp0.4   &  0.6\pp0.1 \\     
          & (f)423\,N & 29\pp18    &  51\pp11   &   18\pp16  &  2\pp2 &  59\pp27 & 38\pp28 &  1.5  &   1.5\pp0.5  &  0.62\pp0.12   &  5.4\pp0.6   &  0.2       \\
	  & 	      & 	   &  	        &  	     &        &          &	   &	   &		  &	           &	          &  	       \\
NGC\,1614 & (j)468\,S & 19\pp31 &  5\pp10 & 60\pp32 & 12\pp31 &  5\pp13 & 82\pp35 & 1.5\pp0.5 & 1.5\pp0.5 & 2.45\pp0.43 &  31\pp10 &  0.2        \\ 
	  & (d)406\,S &  1\pp5  &  8\pp8  & 85\pp14 &  ...    &  2\pp3  & 97\pp10 & 2.0	      & 2.0       & 2.21\pp0.24 &  22\pp8  &  0.2        \\
	  & (c)344\,S & 13\pp11 & 37\pp7  & 48\pp13 &  1\pp1  & 26\pp8  & 72\pp10 & 2.0	      & 2.0       & 2.41\pp0.28 &   8\pp2  &  0.3        \\
	  & (b)281\,S & 43\pp6  & 26\pp4  & 28\pp5  &  2\pp1  & 22\pp5  & 74\pp9  & 2.0	      & 2.0       & 2.77\pp0.28 &   4      &  0.9\pp0.1  \\
	  & (a)219\,S & 13\pp2  & 17\pp2  & 67\pp7  &  ...    &  7\pp1  & 91\pp9  & 2.0	      & 2.0       & 2.52\pp0.25 &   3      &  4.9\pp0.5  \\
	  &    Nuc    &  ...    & 32\pp3  & 64\pp6  &  ...    & 15\pp2  & 84\pp8  & 1.5	      & 2.0       & 1.76\pp0.18 &   3      & 14.0\pp1.4  \\
	  & (e)219\,N & 54\pp6  & 43\pp5  &  ...    & 22\pp4  & 77\pp9  &  ...    & 1.5	      & 1.5       & 1.73\pp0.18 &   2      &  2.0\pp0.2  \\
	  & (f)281\,N & 44\pp5  & 52\pp6  &  ...    & 16\pp3  & 83\pp9  &  ...    & 1.5	      & 1.5	  & 2.36\pp0.24 &   3      &  1.0\pp0.1  \\
	  & (g)344\,N & 63\pp9  & 34\pp7  &  9\pp4  & 13\pp6  & 83\pp15 &  3\pp13 & 1.5	      & 1.5\pp0.5 & 2.71\pp0.29 &   7\pp1  &  0.4\pp0.1  \\ 
 	  & (h)406\,N & 55\pp22 & 26\pp8  &  8\pp18 & 16\pp11 & 65\pp28 & 18\pp33 & 0.5\pp0.5 & 0.5\pp0.5 & 3.06\pp0.38 &  30\pp9  &  0.3        \\
	  & (i)468\,N & 49\pp29 & 48\pp40 &  ...    & 18\pp21 & 81\pp23 &  1\pp3  & 1.0\pp0.5 & 0.5\pp0.5 & 3.93\pp0.41 & 116\pp30 &  0.2\pp0.1  \\
	  & 	      & 	&  	  &  	    &	      &		&	  &	      &		  &	        &	   &  	         \\ 
NGC\,3310 & (d)106\,N &  7\pp15 &  4\pp8 & 94\pp20 & 0\pp1 &  1\pp4  & 97\pp11 & 0.5 & ...	 & 1.53\pp0.24 & 17\pp5 & 0.6\pp0.1 \\	
	  & (c)92\,N  &  0\pp1  & 10\pp3 & 92\pp10 & ...   &  1      & 98\pp10 & 1.0 & 1.0       & 0.66\pp0.07 &  7\pp1 & 0.6\pp0.1 \\ 
	  & (b)78\,N  & 24\pp13 & 19\pp5 & 56\pp14 & ...   &  9\pp6  & 89\pp11 & 1.0 & 1.0       & 0.91\pp0.13 &  5\pp1 & 0.7\pp0.1 \\ 
	  & (a)64\,N  &  6\pp3  & 39\pp5 & 53\pp7  & ...   & 13\pp3  & 86\pp9  & 1.5 & 1.0       & 1.09\pp0.11 &  3     & 1.2\pp0.1 \\	
	  &    Nuc    & 22\pp4  & 22\pp3 & 56\pp7  & 2     &  5\pp2  & 92\pp9  & 1.0 & 1.0       & 1.73\pp0.17 &  2     & 5.0\pp0.5 \\	
	  & (e)64\,S  & 39\pp9  & 47\pp6 & 14\pp11 & 3\pp2 & 48\pp26 & 48\pp27 & 1.0 & 1.0\pp0.5 & 0.39\pp0.07 &  3\pp1 & 1.0\pp0.1 \\  
	  & (f)78\,S  & 35\pp11 & 44\pp6 & 19\pp10 & 3\pp2 & 40\pp19 & 57\pp20 & 0.5 & 0.5       & 0.34\pp0.08 &  4     & 0.7\pp0.1 \\  
	  & (g)92\,S  & 34\pp19 & 43\pp8 & 28\pp17 & 3\pp3 & 39\pp21 & 57\pp23 & 1.0 & 1.0\pp0.5 & 0.27\pp0.08 &  7\pp1 & 0.6\pp0.1 \\  
	  & (h)106\,S & 24\pp13 & 18\pp8 & 57\pp14 & 3\pp3 &  5\pp7  & 91\pp12 & 1.0 & 1.0       & 0.30\pp0.11 & 13\pp4 & 0.6\pp0.1 \\  
	  & 	      & 	&  	 &  	   &       &	     &	       &     &           &	       &	&           \\   
NGC\,7714 & (e)290\,S & 55\pp28 & 11\pp9 & 25\pp26  & 19\pp29 & 22\pp27 & 58\pp43 & 0.5\pp0.5 & 0.5\pp0.5 & 1.49\pp0.31 & 19\pp6  & 0.3       \\
	  & (d)251\,S & 63\pp10 & 34\pp7 &  2\pp6   & 10\pp6  & 81\pp20 &  8\pp20 & 1.0	      & 1.5\pp0.5 & 1.02\pp0.19 &  7\pp2  & 0.3       \\
	  & (c)212\,S & 75\pp8  & 25\pp4 &     ...  & 25\pp6  & 74\pp9  &     ... & 1.0	      & 1.0       & 1.21\pp0.14 &  5\pp1  & 0.8\pp0.1 \\
	  & (b)174\,S & 68\pp7  & 30\pp3 &     ...  & 41\pp5  & 58\pp7  &     ... & 0.5	      & 0.5       & 1.29\pp0.14 &  3      & 1.3\pp0.1 \\
	  & (a)135\,S & 76\pp8  & 23\pp3 &     ...  & 42\pp5  & 56\pp6  &     ... & 0.5	      & 0.5       & 1.12\pp0.12 &  2      & 2.1\pp0.2 \\  
	  &    Nuc    & 59\pp7  & 39\pp5 &     ...  & 15\pp3  & 84\pp9  &     ... & 0.5	      & 0.5       & 1.00\pp0.11 &  2      & 3.2\pp0.4 \\
	  & (f)135\,N & 43\pp6  & 49\pp5 &  6\pp4   &  9\pp2  & 74\pp11 & 16\pp10 & 1.0	      & 1.0       & 1.03\pp0.11 &  2      & 1.6\pp0.2 \\
	  & (g)174\,N & 51\pp8  & 40\pp5 &  8\pp6   &  8\pp3  & 59\pp20 & 32\pp21 & 1.0	      & 1.5	  & 0.88\pp0.10 &  3      & 0.9\pp0.1 \\
	  & (h)212\,N & 34\pp9  & 60\pp9 &  3\pp7   &  3\pp2  & 88\pp18 &  7\pp17 & 1.0	      & 1.0\pp0.5 & 0.44\pp0.10 &  4      & 0.4\pp0.1 \\
	  & (i)251\,N & 69\pp12 & 16\pp9 &  5\pp8   & 20\pp20 & 51\pp30 & 29\pp35 & 0.5	      & 0.5\pp0.5 & 0.45\pp0.14 &  9\pp1  & 0.3       \\
	  & (j)290\,N & 53\pp31 &  4\pp8 & 29\pp30  & 24\pp37 & 10\pp23 & 65\pp44 & 0.5	      & 0.5\pp0.5 & 1.51\pp0.32 & 28\pp9  & 0.3       \\
\noalign{\smallskip}	     	 																	    
\hline																					  
\noalign{\smallskip}
\end{tabular}
\begin{list}{Table Notes:}
\item (1), (2), (3): average contribution in flux of the \textit{young} ($\rm t \leq 50\times10^6$ yr), \textit{intermediate-age} ($\rm 50\times10^6 < t \leq 2\times10^9$ yr) 
and \textit{old} ($\rm t > 2\times10^9$ yr)
SP component, respectively; (4), (5), (6): average contribution of the SP components in mass; (7), (8): flux- and mass-weighted mean metallicities; (9) visual extinction;
(10) percent mean deviation $\rm |O_{\lambda} - M_{\lambda}|/O_{\lambda}$, where $\rm O_{\lambda}$ is the observed spectrum and $\rm M_{\lambda}$ is the fitted model and (11) reduced chi square.
Whenever there are no error value, the uncertainties are zero.
\end{list}
\end{table*}

Below we describe the results obtained through our SP analysis for each galaxy and compare them with the results found in the literature.

\subsection{NGC\,34}\label{34}

Panels {\it a} and {\it b} of Fig.~\ref{hist34} show a predominance of younger ages in the nuclear region shifted towards the south direction, 
while the intermediate age SP is more pronounced north of the nucleus. 
The contribution of the old SP is enhanced in the mass fraction histogram as expected,
once the older the SP is, less light it will radiate, decreasing the flux contribution and increasing the 
mass contribution of this age in the SP components. These results are in agreement with the literature. For example, 
studying the optical properties of NGC\,34 (on a larger field of view) \citet{sch07} conclude 
that this source supports a rich system of young massive star clusters, a blue exponential disk and 
a strong gaseous outflow, all signatures of a recent gas-rich merger accompanied by a strong starburst.
This source was also studied by R08 in the NIR part of the spectrum. analyzing
the 230 central parsecs of this galaxy, they found a dominating intermediate age SP ($\sim$ 1Gyr).

In Fig.~\ref{hist34}{\it c} we show the mass and flux-weighted mean metallicities along the apertures of NGC\,34. 
Our results point to above-solar mean metallicity values along the galaxy.
Moreover, the apertures displaying higher mass-weighted mean metallicity values than the flux-weighted ones, are those
in which the contribution of the old SP increases. Such discrepancy can be associated to the well known age-metallicity degeneracy, i.e. 
for a fixed mass, a high-metallicity SP looks cooler - and older - than a low-metallicity SP, thus resulting in a higher 
$M/L$ ratio. In this context, one can interpret that the flux-weighted metallicity is more sensitive to the 
young component, while the mass-weighted metallicity is more sensitive to the old one.

The north side of the galaxy display lower visual extinction ($A\rm_{v} <$2.0\,mag) 
than the nuclear ($A\rm_{v} \sim$3.0\,mag) and southern region ($A\rm_{v} \sim$2.0\,mag), suggesting a differential reddening.
To better compare our results with those found in in the literature, we estimated a weighted average of the visual extinction values derived
by {\sc starlight}, resulting in $\bar{A}\rm_{v}$=2.49\pp0.10\,mag for NGC\,34. Using the values of {\it E(B-V)} quoted by \citet{veil95}, 
\citet{gol97a} derived $A\rm_{k}$=0.67\,mag (in a 1$\arcsec$.5 $\times$ 4$\arcsec$.5 beam), which results in $A\rm_{v}$=6.2\,mag 
when using the relation $A\rm_{k}$=0.108$A\rm_{v}$ \citep{mathis90}.

In order to assess the robustness of the fit, we display the ${\chi}^2$ and Adev (percent mean deviation) along the apertures
in Fig.~\ref{hist34}{\it e}, the continuum profile at ${\lambda}_{cent}=12230\rm{\AA} $ in Fig.~\ref{hist34}{\it f} and the SNR along the galaxy in Fig.~\ref{hist34}{\it g}.
As the SNR drops to values below 10, the percent mean deviation (Adev) reaches higher values (see Figs~\ref{hist1614}{\it e} to {\it g}).

\subsection{NGC\,1614}\label{1614}

From Fig.~\ref{hist1614}{\it a} we can see a peak of old SP in nucleus at 219\,pc south, while the off-nuclear northern apertures present 
a dominant young/intermediate age SP component. The fact that we are not detecting a contribution of the young SP ($t \leq 50\times10^6$ yr) in the
nuclear aperture can be explained in two different scenarios: (i) the presence of an AGN, which would quench star formation in the nuclear region \citep{nesv06}, or 
(ii) that the NIR still cannot reveal the young stars embedded in dustier nuclear regions \citep{imani13}. By modeling the starburst of this galaxy, \citet{alo01} associated the strong CO 
concentrations found in the nucleus with the a older part of the starburst ($10^9$ yr old or older stars), 
which is similar to the dominant SP age we found in the nuclear aperture. 

Moreover, we found evidence of a circumnuclear ring-like structure \footnote{As we used an unidimensional slit, we can only see an increase in the young/
intermediate age SP in both sides of the nucleus, in this sense we cannot assert if this pattern surrounds the nuclear region. Although, from 
\citet[][ - Fig.4]{alo01} we can see that the Pa$\alpha$ emission image reveals a nuclear ring-like structure with approximate diameter of 650\,pc, possibly formed by H{\sc ii} regions. 
A deeper analysis will be made using Integral Field Unit (IFU) data of this source.} of young/intermediate age SP with a diameter of about 600\,pc.
In fact, \citet{olss10} detected a ring-like structure traced out by 1.4\,GHz and 5\,GHz MERLIN contours with a radius of around 310\,pc, which
is consistent with our results. This also has been reported by \citet{alo01},
which have stated that the relation between the strong stellar CO bands to surrounding ionized gas rings to molecular gas suggests that 
the luminous starburst started in the nucleus and is propagating outward into a molecular ring. 
\citet{vaisanen12}, through the analysis of L-band IFU observations of the inner Kpc of NGC\,1614, detected a ring that extends from 200 to 500\,pc from the nucleus,
from the equivalent width map of 3.3$\mu$m polycyclic aromatic hydrocarbon (PAH) emission. It is worth mentioning that such kind of nuclear 
rings are quite common in AGNs, but rarely found in these systems \citep[e.g.][]{riffel11b,riffel10,thaisa12}.

The increase in the young/intermediate age SP component at $\sim$300\,pc north from the nucleus can also be interpreted as being an evidence of
the secondary nucleus (see Fig.~\ref{hist1614}{\it a}), also reported by \citet{alo01} as the remnant of the merger with a smaller galaxy,which has by now
largely been destroyed. In a previous work, R08 found a 1\,Gyr old SP dominating the light in the the inner 154\,pc of this source.

We found above solar values for the flux and mass-weighted mean metallicity along the galaxy (see Fig.~\ref{hist1614}{\it c}).
As the contribution of the young SP component grows north, the metallicity tends to decrease, suggesting a metal-poor young SP at this side of the galaxy.
In this scenario, the modest sized galaxy which merged with NGC\,1614 in the past \citep{alo01} would have diluted the gas in the remnant, 
thus giving rise to a metal-poor SP, poorer than the old SP component already present in NGC\,1614.

The values derived by us for the extinction ($A\rm_{v}$) range from 1.7 to 3.9\,mag ($\bar{A}\rm_{v}$=2.66\pp0.08). 
Results found in the literature are in agreement with ours, for example \citet{neff90} derived $A\rm_{v}$=3-5\,mag 
assuming a foreground dust screen model, from optical and infrared colors (in a $\sim$20$\arcsec \times$ 20$\arcsec$ beam), 
as well as infrared emission lines (in a 2$\arcsec$.7 $\times$ 4$\arcsec$.3 beam). Using NIR colors, \citep{srr96} found $A\rm_{v}$=4.9\,mag 
(in a 2$\arcsec$.4 $\times$ 4$\arcsec$.8 beam), while \citet{kot01} (in a 2$\arcsec \times$ 2$\arcsec$ beam) derived $A\rm_{v}$=1.8\,mag southeast of the nucleus and $A\rm_{v}$=3.5\,mag northwest.
Assuming an intrinsic stellar color {\it H-K}=0.2, \citet{oliva95} derived $A\rm_{v} \sim$4\,mag (in a 2$\arcsec$.2 $\times$ 4$\arcsec$.4 beam). 
Moreover, \citet{alo01}, using the flux of two [Fe{\sc ii}] emission 
lines at 1.257 and 1.644$\mu$m as well as the hydrogen recombination lines Pa$\beta$ and Br$\gamma$ and a foreground dust screen model, 
derived the extinction to the gas in the K-band as $A\rm_{k}$=0.40-0.49\,mag (in a 19$\arcsec$.5 $\times$ 19$\arcsec$.5 beam), corresponding to $A\rm_{v} \sim$4\,mag, 
using \citet{mathis90} relation of $A\rm_{v}$ and $A\rm_{k}$. Also using recombination lines, \citet{kot01} found $A\rm_{v}$=3.8\,mag 
(in a 3$\arcsec$.5 $\times$ 3$\arcsec$.5 beam), a similar value than the one \citet{bush86} derived ($A\rm_{v} \sim$3\,mag) for a similar beam.
Likewise, \citet{gol95} derived $A\rm_{v}$=2.96\,mag (in 3$\arcsec \times$ 12$\arcsec$ beam), using the values of {\it E(B-V)} quoted by \citet{veil95}.

On the other hand, \citet{pux94} present discrepant values when compared to those we derived and those found in the literature.
Using hydrogen recombination line fluxes they inferred a total visual extinction of $A\rm_{v}$=15\pp2.5\,mag (in 9$\arcsec \times$ 3$\arcsec$ beam) assuming a 
composite model (a mixture of dust, gas and foreground screen) and claimed that the extinction on NGC\,1614 could not be modeled with a simple foreground 
dust screen model, for which they found visual extinction values ranging from 4.7 to 9.5\,mag.

\subsection{NGC\,3310}\label{3310}

Our results point out to a predominant young/intermediate age SP southwards (see Fig.~\ref{hist3310}{\it a}). 
An increase in the older component can be seen in both flux and mass contribution towards the opposite direction. 
Similar to our study, several previous works in the optical region \citep{balick81,tega84,sch88} estimated the
starburst age on this source in the range of $10^7$ to $10^8$ yr.

The nucleus and six surrounding H{\sc ii} regions, four of which are located 
at less than 400\,pc from the galaxy nucleus (a wider range than the one analyzed in this work, which comprehend only
the nuclear region studied by these authors\footnote{It it worth mentioning that the apertures used in this source are the most inner ones in the sample, 
ranging only up to 106\,pc from the nucleus.}) were investigated by \citet{mgp93} between 3600 and 9600\,$\rm{\AA} $.
In this study, they propose a two-age model (5 and 15\,Myr) for the starburst activity and low metallicity values (0.2 - 0.4\,Z$_{\odot}$) for the circumnuclear 
(400\,pc from the centre) and disc H{\sc ii} regions, while the nucleus presented solar abundances, in agreement with our 
results (Z$\sim$ Z$_{\odot}$ along the galaxy - see Fig.~\ref{hist3310}{\it c}). R08 also studied the inner 56\,pc of NGC\,3310 
in the NIR part of the spectrum, detecting a dominant 1\,Gyr old SP. Similar to the scenario we propose for 
NGC\,1614, the low metallicity SP derived for this source suggests that the gas present in the galaxy interacting with NGC\,3310
\citep{balick81} should be metal poor, given rise to a metal poorer SP. 

We found a extinction of $A\rm_{v} \sim$1.7\,mag in the nucleus, ranging from 0.27\,mag southwards to 1.5\,mag toward the 
north direction ($\bar{A}\rm_{v}$=1.43\pp0.06). As for NGC\,1614, \citet{gol95} derived the extinction for this source and 
found $A\rm_{v}$=2.69\,mag (in 3$\arcsec \times$ 9$\arcsec$ beam), a slightly larger value than those we found. 
It is worth mentioning that this is the less interacting galaxy of the sample.

\subsection{NGC\,7714}\label{7714}

The analysis of the flux contribution of the SP vectors from Fig.~\ref{hist7714}{\it a} clearly shows the predominance of the 
young SP component along the galaxy. While the contribution of this young SP component increases toward southeast, 
the intermediate age SP component increases in the opposite direction. \citet{bern93} has presented an evolutionary model for 
the pair NGC\,7714 and NGC\,7715 and stated that NGC\,7715 is in the postburst phase, starting its star formation around 10$^8$ yr ago, similar to our results. 
A tidal tail between these two galaxies can bee seen in \citet{smwa92}, and we suggest that the SP distribution northward 
(increasing intermediate age SP component) can be related to this poststarburst population in NGC\,7715. In fact, such 
an age is consistent with the scenario proposed by \citet{rosa99}, where the presence of red super giants is associated with the 
Ca {\sc ii} triplet, present in the galaxy spectrum.  Besides, this result is further supported by the fact that this is the 
side in which NGC\,7714 interacted with NGC\,7715 in the past \citep{ss97,ss03}. Several previous studies in the UV and optical 
part of the spectrum \citep{rosa95,rosa99,lancon01,cid03} detected young age SP from the inner hundreds to thousands of 
parsecs, in agreement with our work. R08 also found a dominant star formation burst with 1\,Gyr old studying the inner 
115\,pc of this source.  

The metallicity, as can be seen from Fig.~\ref{hist7714}{\it c}, increases with the increase of the intermediate SP, 
as expected. The lower mean metallicity values derived for this source can be explained in the scenario proposed by \citet{rosa95}, 
in which the unprocessed gas from the companion NGC\,7715 could be the source of fuel to the starburst in NGC\,7714.

As we can see from Fig.~\ref{hist7714}{\it d}, the visual extinction is nearly constant along the galaxy with $\bar{A}\rm_{v}$=1.13\pp0.04, 
similar to the results found in the literature. For example, \citet{kot01} derived $A\rm_{v}$=1.2\,mag (in a 6$\arcsec \times$ 7$\arcsec$ beam) 
from NIR colors and ${A}\rm_{v}$=1.7\,mag from recombination line fluxes, while \citet{oliva95} found $A\rm_{v}$1-2\,mag (in a 2$\arcsec$.2 $\times$ 4$\arcsec$.4 beam)
assuming an intrinsic stellar color {\it H-K}=0.2. Moreover, \citet{pux94} found $A\rm_{v}$=1.8\pp0.7\,mag for a point-source model or $A\rm_{v}$=3.9\pp1.7\,mag 
if sources and dust are homogeneously distributed. They claim, however, that different from NGC\,1614, NGC\,7714 presents low extinction values and can 
indeed be modeled with a simple foreground dust screen model.


\subsection{Emission Gas}

The sample of galaxies analyzed in this work display strong emission lines in their spectra, as we can see from Figs~\ref{especs34} to \ref{especs7714}. 
Apart from the absorption features that shape the continuum emission, these sources have been classified as starforming/H{\sc ii} galaxies by 
several previous works, particularly in the optical region \citep[e.g.][]{rosa99,alo01,wehgal06,sch07}. 
Thus,  we measured these nebular atomic emission lines ([S{\sc iii}]\,$\lambda$9530, He{\sc i}\,$\lambda$10830, [Fe{\sc ii}]\,$\lambda$12570, 
Pa$\beta$\,$\lambda$12810, [Fe{\sc ii}]\,$\lambda$16440 and  Br$\gamma$\,$\lambda$21650) and the H$_2$\,$\lambda$21210 line. It is important to mention 
that the standard procedure to properly measure emission lines consist of subtracting the stellar continuum fitted by stellar population 
synthesis methods. However, we did not subtract the stellar continuum, because the emission lines
were severely diluted as a result of the smoothing done in the observed spectra to match the resolution of M05 models (R$\leq$ 250 - see Sec.~\ref{spsm}).

The emission-line fluxes can be used, for example, to estimate the reddening effects on the emission gas. Therefore, we derived the interstellar extinction {\it c}, 
the color excess {\it E(B-V)} and the visual extinction $A\rm_{v}$ (Eq.~\ref{av}) for each aperture of the galaxies by means of the emission lines. 
To this purpose we assume the ratio of total to selective extinction as R{\sc v}=4.05\pp0.80, from \citet{calzetti00}, which is the most suitable one when dealing with SBs 
\citep{calzetti00,fisch03}.
Moreover, we adopt the intrinsic value of 5.86 for the emission-line ratio Pa$\beta$/Br$\gamma$, following the 
case B \citep[T=10\,000K, N=10$^2$cm$^{-3}$][]{osterbrock}. Using the above values one can determine $A\rm_{v}$ following the equation:

\begin{equation}
\label{av}
\rm A_{v}=-15.24log\left(\frac{1}{5.88} \frac{Pa\beta}{Br\gamma} \right)
\end{equation}

\noindent
The measured fluxes are listed in Tab.~\ref{flux} and the reddening parameters in Tab.~\ref{flux_ind}.
In a previous work, as we have already mentioned, R08 performed SP synthesis in the nuclear region of the four galaxies studied here. They derived lower values 
for the color excess {\it E(B-V)} than those present in Tab.~\ref{flux_ind}. We suggest that this difference can be related to the use of CCM extinction law by 
R08 when fitting the stellar population, which is not the most suitable on for SBs \citep{calzetti94,fisch03}. 

In fact, when comparing the results for $A\rm_{v}$ obtained with the emission-line ratio Pa$\beta$/Br$\gamma$ (Tab.~\ref{flux_ind} - Col.2) 
and those derived by {\sc starlight} code (Tab.~\ref{flux_ind} - Col.3), we can see that the mean values derived from the emission lines are 
larger than those obtained from the spectral fitting.  A possible explanation for this discrepancy is to consider that the hot ionizing 
stars could be associated to a dustier region with respect to the cold stellar population \citep{calzetti94}. Another important aspect is that
as the merger experienced by the galaxies progresses, tidally induced gas motions and outflows from galactic winds become more frequent\citep[e.g.][]{heck90}.
In this scenario, shocks induced by large-scale gas flows have an influence on the emission line gas \citep[e.g.][]{colina05} and can contaminate the line ratios.

The emission lines can also be used to estimate the SFR of the sources. Adopting SFR=L(H$\alpha$)$\times7.9\times10^{-42}$ and L(H$\alpha$)=103L(Br$\gamma$) \citep{kenn98},  
we derived the SFR for each aperture of our galaxy sample using their reddening corrected Br$\gamma$ luminosities. The SFRs are listed in Tab.~\ref{flux_ind}. 
Since {\sc starlight} outputs the mass that have been processed into stars over the last {\it t} years (M$_\star^{t}$), this can be used to estimate the mean SFR over 
a period of time {\it t}. We have estimated the mean SFR$\star$ over the last {\it t}$\leq$10\,Myrs as being the ratio of M$_\star ^{t}$/{\it t}. The 
obtained values are also listed in Tab.~\ref{flux_ind}.

As we can see from Tab.~\ref{flux_ind}, the nuclear SFRs derived through Br$\gamma$ (Col.5) are those that exhibit the major discrepancies when compared to 
those calculated through {\sc starlight} output. In fact, since our sample comprises only SBs, it is likely that a significant fraction of the star formation
occurs in dusty regions \citep{imani13}. In this sense, our results of the SFR may suggest that the NIR spectral range would probe weaker-obscured star-forming
regions. Thus, we may miss dustier starburst regions, once our values tend to be smaller than those calculated in larger wavelengths.

For NGC\,34, \citet{ximena10} calculated the SFR from radio luminosity and found a value of 64\,M$\odot$yr$^{-1}$ 
(in a 3$\arcmin \times$ 2$\arcmin$ beam), in agreement with SFR$\sim$50\,M$\odot$yr$^{-1}$ derived by \citet{prouton04} from the FIR luminosity for similar beam.
For NGC\,1614, \citet{alo01} found SFR=52\,M$\odot$yr$^{-1}$ from integrated FIR luminosity (in a 45$\arcsec \times$ 250$\arcsec$ beam) and assuming that 70\%
of this star formation occurs in the nucleus, they predict a nuclear SFR of 36\,M$\odot$yr$^{-1}$.  
For NGC\,1614 and NGC\,7714, assuming a constant star formation model, \citet{kot01} derived SFRs ranging from 2.1-4.8\,M$\odot$yr$^{-1}$ (in a 3.9$\arcsec$ aperture)
and 0.1-3.9\,M$\odot$yr$^{-1}$ (in a 11$\arcsec$.2 aperture), respectively, while assuming a model of an instantaneous burst of star formation (ISF) they derived SFRs ranging from 
11-25\,M$\odot$yr$^{-1}$ and 0.4-16\,M$\odot$yr$^{-1}$, respectively. The results derived assuming a ISF are closer to the values derived by us for NGC\,1614 and NGC\, 7714 and we believe 
that the reason for that is the resembling spacial scale used in their work in relation to this study. 
Moreover, we suggest that the negligible value derived for the nuclear SFR of NGC\,1614 by {\sc starlight} can be explained if we consider the LINER classification of this source.
In this scenario, the strong emission lines present in the nuclear spectra of this source can be associated to an AGN-activity, which would quench the star formation \citep{nesv06} 
in the nucleus and limited it to the outer regions\footnote{A proper analysis of this evidence would require an IFU study of the nuclear region of this galaxy.}. 
It is important to reinforce the scenario in which shocks induced by large-scale gas flows in merging systems affect the emission line gas and may contaminate the line ratios
used to determine the SFRs and $A\rm_{v}$ for example.

\begin{table*}
\centering
\scriptsize
\caption{Measured fluxes in units of 10$^{-15}$\,erg\,cm$^{-2}$\,s$^{-1}$.}
\label{flux}
\begin{tabular}{lllllllll}
\noalign{\smallskip}
\hline
\hline
\noalign{\smallskip}
 Galaxy  & Aperture & [S{\sc iii}]    & He{\sc i}      & [Fe{\sc ii}]	   & Pa$\beta$      & [Fe{\sc ii}]    & H$_2$ & Br$\gamma$    \\
 	 &	    & 0.953$\mu$m     & 1.083$\mu$m    & 1.257$\mu$m       & 1.281$\mu$m    & 1.644$\mu$m     & 2.121$\mu$m        & 2.165$\mu$m   \\
 	 &	    & (1) 	    &     (2)  		  &    (3)       &       (4) 	   &      (5) 	    & 		(6)        & (7)	   \\
\noalign{\smallskip}
\hline
\hline
\noalign{\smallskip}  
NGC\,34   & (c)423\,S &  0.54\pp0.21 &  0.46\pp0.17 & ...	  & 0.23\pp0.08  & 0.25\pp0.14 & 0.25\pp0.03  & 0.09\pp0.03 \\  
          & (b)346\,S &  1.15\pp0.31 &  0.65\pp0.26 & 0.38\pp0.12 & 0.37\pp0.13  & 0.00\pp0.00 & 0.73\pp0.11  & 0.22\pp0.06 \\
          & (a)269\,S &  1.46\pp0.30 &  2.24\pp0.52 & 1.63\pp0.15 & 1.23\pp0.11  & 1.78\pp0.45 & 2.08\pp0.22  & 0.81\pp0.12 \\ 
          &    Nuc    & 10.60\pp0.84 & 12.00\pp1.66 & 8.41\pp0.82 & 7.58\pp0.71  & 9.32\pp2.34 & 8.21\pp0.75  & 4.45\pp0.54 \\
          & (d)269\,N &  2.77\pp0.32 &  2.92\pp0.59 & 1.74\pp0.17 & 1.06\pp0.11  & 1.65\pp0.48 & 1.48\pp0.12  & 0.68\pp0.08 \\
          & (e)346\,N &  0.70\pp0.21 & ...	    & 0.48\pp0.14 & 0.38\pp0.11  & ...         & 0.63\pp0.07  & ...	    \\
          & (f)423\,N &  ...	     & ...	    & ...	  & ...	    	 & ...	       & ...	      & ...	    \\
	  &	      &		     &		    &		  &		 &	       &	      &	   	    \\ 
NGC\,1614 & (j)468\,S &   0.80\pp0.19 &   1.13\pp0.20 &   0.41\pp0.15 &  0.57\pp0.09 &  0.28\pp0.06 & 0.08\pp0.03 &   0.15\pp0.04  \\	  
	  & (d)406\,S &   1.55\pp0.28 &   1.89\pp0.26 &   0.50\pp0.08 &  0.92\pp0.08 &  0.74\pp0.08 & 0.14\pp0.03 &   0.36\pp0.03  \\
	  & (c)344\,S &   4.14\pp0.36 &   3.20\pp0.23 &   1.12\pp0.09 &  2.68\pp0.10 &  1.28\pp0.11 & 0.53\pp0.06 &   1.12\pp0.07  \\	  
	  & (b)281\,S &  14.40\pp0.36 &  11.00\pp0.32 &   2.78\pp0.18 &  9.46\pp0.17 &  3.33\pp0.24 & 0.83\pp0.09 &   3.47\pp0.11  \\ 
	  & (a)219\,S &  46.20\pp0.44 &  33.90\pp0.64 &   7.11\pp0.27 & 25.80\pp0.27 &  8.17\pp0.84 & 1.36\pp0.17 &   9.67\pp0.20  \\  
	  &    Nuc    & 227.00\pp1.79 & 187.00\pp3.60 &  28.20\pp1.49 & 97.30\pp1.32 & 17.80\pp4.40 &   ...	  &  32.00\pp1.68  \\
	  & (e)219\,N &  64.40\pp0.33 &  51.00\pp0.59 &   7.79\pp0.34 & 29.80\pp0.32 &  8.84\pp0.85 & 1.35\pp0.15 &  10.40\pp0.17  \\ 
	  & (f)281\,N &  27.90\pp0.35 &  21.30\pp0.23 &   4.12\pp0.17 & 13.90\pp0.16 &  4.61\pp0.24 & 0.98\pp0.08 &   4.63\pp0.08  \\ 
	  & (g)344\,N &   9.61\pp0.31 &   7.01\pp0.22 &   1.48\pp0.14 &  5.21\pp0.14 &  1.93\pp0.11 & 0.65\pp0.07 &   1.71\pp0.07  \\	 
	  & (h)406\,N &   3.37\pp0.31 &   2.24\pp0.25 &   0.57\pp0.10 &  1.87\pp0.10 &  0.57\pp0.09 & 0.43\pp0.07 &   0.65\pp0.07  \\	 
	  & (i)468\,N &   0.65\pp0.17 &   0.91\pp0.21 &   0.36\pp0.09 &  0.73\pp0.10 &    ...	    & 0.24\pp0.07 &   0.13\pp0.05  \\	 
	  &	      &		      &		      &		      &		     &		    &		  &	           \\					
NGC\,3310 & (d)106\,N & ...         & ...         & ...	        & ...		& ...	      & ...	     & ...	   \\
	  & (c)92\,N  & 0.09\pp0.03 & ...	  & ...	        & 0.03\pp0.02	& ...	      & ...	     & ...	   \\
	  & (b)78\,N  & 0.26\pp0.06 & 0.13\pp0.04 & ... 	& ...	       	& ...	      & ...	     & 0.03\pp0.01 \\
	  & (a)64\,N  & 0.73\pp0.04 & 0.40\pp0.03 & 0.07\pp0.02 & 0.28\pp0.02	& 0.13\pp0.03 & ...	     & 0.10\pp0.01 \\
	  &    Nuc    & 6.58\pp0.10 & 3.57\pp0.11 & 0.75\pp0.07 & 2.44\pp0.07	& 0.86\pp0.13 & ...	     & 0.77\pp0.05 \\
	  & (e)64\,S  & 0.74\pp0.05 & 0.38\pp0.05 & 0.09\pp0.02 & 0.19\pp0.02	& 0.11\pp0.03 & 0.06\pp0.02  & 0.06\pp0.01 \\
	  & (f)78\,S  & 0.31\pp0.06 & 0.25\pp0.08 & ... 	& 0.07\pp0.03	& 0.05\pp0.02 & ...	     & ...	   \\
	  & (g)92\,S  & 0.15\pp0.06 & ...	  & ...		& ...  	  	& ... 	      & ...	     & ...    	   \\
	  & (h)106\,S & ...         & ...         & ...         & ...           & ...         & ... 	     & ...	   \\
	  &	      &		    &		  &		&		&	      &	    	     &  	   \\     
NGC\,7714 & (e)290\,S &   3.61\pp0.25 &   1.82\pp0.19 &  0.36\pp0.08 &   1.21\pp0.07 &  0.33\pp0.08 & 0.10\pp0.04 &  0.30\pp0.05 \\
	  & (d)251\,S &   7.17\pp0.37 &   4.27\pp0.18 &  0.48\pp0.14 &   2.58\pp0.14 &  0.69\pp0.13 & ...	  &  0.47\pp0.05 \\
	  & (c)212\,S &  15.90\pp0.27 &  10.10\pp0.17 &  0.66\pp0.16 &   5.24\pp0.12 &  0.71\pp0.13 & 0.17\pp0.05 &  1.14\pp0.05 \\
	  & (b)174\,S &  37.30\pp0.33 &  23.50\pp0.25 &  1.98\pp0.16 &  11.20\pp0.13 &  1.96\pp0.21 & ...	  &  2.78\pp0.08 \\
	  & (a)135\,S &  59.90\pp0.40 &  41.10\pp0.41 &  2.98\pp0.20 &  18.50\pp0.17 &  2.59\pp0.27 & ...	  &  4.55\pp0.11 \\
	  &    Nuc    & 195.00\pp1.89 & 132.00\pp2.40 & 17.00\pp0.91 &  63.30\pp0.97 & 17.10\pp2.83 & 2.56\pp0.63 & 15.90\pp0.66 \\
	  & (f)135\,N &  24.40\pp0.28 &  14.90\pp0.33 &  4.06\pp0.17 &   9.12\pp0.15 &  3.76\pp0.34 & 0.55\pp0.08 &  2.51\pp0.09 \\
	  & (g)174\,N &  14.60\pp0.20 &   8.67\pp0.20 &  2.43\pp0.16 &   5.69\pp0.13 &  2.24\pp0.17 & 0.39\pp0.07 &  1.44\pp0.06 \\
	  & (h)212\,N &   5.40\pp0.21 &   3.21\pp0.14 &  0.85\pp0.08 &   2.14\pp0.08 &  1.00\pp0.08 & 0.19\pp0.06 &  0.57\pp0.04 \\
	  & (i)251\,N &   1.27\pp0.18 &   0.98\pp0.14 &  0.29\pp0.07 &   0.74\pp0.08 &  0.29\pp0.07 & ...	  &  0.09\pp0.03 \\		
	  & (j)290\,N &   0.25\pp0.10 &   0.14\pp0.07 &  ...	     &   0.14\pp0.04 &  0.14\pp0.06 & ...	  & ... 	 \\
\noalign{\smallskip}																			   
\hline																					  
\noalign{\smallskip}
\end{tabular}
\end{table*}

\begin{table*}
\centering
\scriptsize
\caption{Emission-line Ratio Pa$\beta$/Br$\gamma$ derived quantities: {\it c}, $A\rm_{v}$, {\it E(B-V)} and SFR.}
\label{flux_ind}
\begin{tabular}{llllllll}
\noalign{\smallskip}
\hline
\hline
\noalign{\smallskip}
 Galaxy  & Aperture & {\it c} &  $A\rm_{v}$\,(mag) &  ${A\rm_{v}}^{\star}$\,(mag)  & {\it E(B-V)}\,(mag)   & SFR\,M$\odot$yr$^{-1}$  & SFR$^{\star}M\odot$yr$^{-1}$ \\
 	 &	    & (1)     &  (2) &      (3)   &  (4)    &(5) & (6) \\
\noalign{\smallskip}  
\hline
\hline
\noalign{\smallskip}  
NGC\,34   & (c)423\,S & 2.21\pp1.28 & 5.52\pp0.32  &   2.86\pp0.29   &  1.36\pp0.22 & 0.13\pp0.08 &  3.65\pp0.60 \\  
          & (b)346\,S & 3.32\pp1.18 & 8.28\pp0.29  &   2.13\pp0.22   &  2.05\pp0.33 & 0.48\pp0.26 &  0.65\pp0.21 \\  
          & (a)269\,S & 3.59\pp0.46 & 8.96\pp0.11  &   1.79\pp0.19   &  2.21\pp0.35 & 1.97\pp0.46 &  0.02\pp0.06 \\  
          &    Nuc    & 3.28\pp0.41 & 8.20\pp0.10  &   3.23\pp0.32   &  2.02\pp0.32 & 9.63\pp1.93 & 34.64\pp3.68 \\  
          & (d)269\,N & 3.52\pp0.42 & 8.79\pp0.10  &   1.78\pp0.18   &  2.17\pp0.34 & 1.62\pp0.32 &  2.43\pp0.65 \\  
          & (e)346\,N &   ...	    &...	   &   0.97\pp0.10   &  ...	    &	...	  &  0.28\pp0.10 \\	      
          & (f)423\,N &   ...	    &...	   &   0.62\pp0.12   &  ...	    &	...	  &  0.20\pp0.18 \\	
          & Mean      & 2.99\pp0.24 & 7.45\pp0.06  &   2.49\pp0.10   &  2.01\pp0.13 &	          &      \\          
	  &	      &		    &              &                 &              &             &             \\
NGC\,1614 & (j)468\,S &  1.16\pp0.82   & 2.89\pp0.20 &   2.52\pp0.25  &  0.71\pp0.12 &  0.09\pp0.04  &  1.11\pp0.14 \\  
	  & (d)406\,S &  2.21\pp0.32   & 5.52\pp0.08 &   2.77\pp0.28  &  1.36\pp0.22 &  0.34\pp0.05  &  1.28\pp0.20 \\  
	  & (c)344\,S &  2.38\pp0.19   & 5.95\pp0.05 &   2.41\pp0.28  &  1.47\pp0.23 &  1.13\pp0.11  &  0.18\pp0.16 \\  
	  & (b)281\,S &  2.04\pp0.10   & 5.09\pp0.02 &   2.21\pp0.24  &  1.26\pp0.20 &  3.04\pp0.15  &  0.00\pp0.03 \\  
	  & (a)219\,S &  2.09\pp0.06   & 5.23\pp0.02 &   2.45\pp0.43  &  1.29\pp0.20 &  8.67\pp0.28  &  0.08\pp0.14 \\  
	  &    Nuc    &  1.75\pp0.14   & 4.36\pp0.04 &   1.76\pp0.18  &  1.08\pp0.17 & 25.06\pp1.93  &  ...	  \\  
	  & (e)219\,N &  1.91\pp0.05   & 4.76\pp0.01 &   1.73\pp0.18  &  1.17\pp0.19 &  8.66\pp0.23  & 12.44\pp1.9  \\  
	  & (f)281\,N &  1.78\pp0.06   & 4.45\pp0.01 &   2.36\pp0.24  &  1.10\pp0.17 &  3.67\pp0.10  &  6.60\pp1.43  \\  
	  & (g)344\,N &  1.74\pp0.13   & 4.35\pp0.03 &   2.71\pp0.29  &  1.07\pp0.17 &  1.34\pp0.09  &  0.89\pp0.15  \\  
	  & (h)406\,N &  1.89\pp0.32   & 4.73\pp0.08 &   3.06\pp0.38  &  1.17\pp0.18 &  0.54\pp0.09  &  0.24\pp0.14  \\  
	  & (i)468\,N &  0.12\pp1.08   & 0.30\pp0.27 &   3.93\pp0.41  &  0.08\pp0.02 &  0.06\pp0.03  &  0.04\pp0.03  \\  
	  & Mean      &  1.17\pp0.03   & 2.94\pp0.01 &   2.66\pp0.08  &  1.19\pp0.02 &              &       \\
	  &	      &		       &	     &	      	     & 	       	    &	            &	           \\ 
NGC\,3310 & (d)106\,N &  ...          &   ...	    &  1.09\pp0.11 &   ...	     &   ...	     & ...	  \\ 
	  & (c)92\,N  &  ...          &   ...	    &  0.91\pp0.13 &   ...	     &   ...	     & ...	  \\ 
	  & (b)78\,N  &  ...          &   ...	    &  0.66\pp0.07 &   ...	     &   ...	     &  ...	  \\ 
	  & (a)64\,N  &  1.97\pp0.33  & 4.91\pp0.08 &  1.53\pp0.24 &   1.21\pp0.19   & ...	     & ...	  \\ 
	  &    Nuc    &  1.64\pp0.19  & 4.09\pp0.05 &  1.73\pp0.17 &   1.01\pp0.16   & 0.02	     & 0.12\pp0.02 \\ 
	  & (e)64\,S  &  1.64\pp0.52  & 4.10\pp0.13 &  0.39\pp0.07 &   1.01\pp0.16   & ...	     &  ...	  \\ 
	  & (f)78\,S  &  ...          &   ...	    &  0.34\pp0.08 &   ...	     &   ...	     &  ...	  \\ 
	  & (g)92\,S  &  ...          &   ...	    &  0.27\pp0.08 &   ...	     &   ...	     &  ...	  \\ 
	  & (h)106\,S &  ...          &   ...	    &  0.30\pp0.11 &   ...	     &   ...	     &  ...	  \\ 
	  & Mean      &  1.74\pp0.16  & 4.35\pp0.04 &  1.43\pp0.06 &   1.08\pp0.01   &               &     	  \\
	  &	      &		      &	            &	           &	             &  	     &            \\     
NGC\,7714 & (e)290\,S &  1.00\pp0.47  & 2.49\pp0.12 &  1.12\pp0.12  &  0.62\pp0.10  & 0.06\pp0.01   &  2.58\pp0.31   \\  
	  & (d)251\,S &  0.18\pp0.32  & 0.45\pp0.08 &  1.29\pp0.14  &  0.11\pp0.02  & 0.07\pp0.01   &  1.63\pp0.24   \\  
	  & (c)212\,S &  0.65\pp0.13  & 1.63\pp0.03 &  1.21\pp0.14  &  0.40\pp0.06  & 0.20\pp0.01   &  0.36\pp0.09   \\  
	  & (b)174\,S &  1.00\pp0.08  & 2.50\pp0.02 &  1.02\pp0.19  &  0.62\pp0.10  & 0.56\pp0.02   &  0.15\pp0.05   \\  
	  & (a)135\,S &  0.98\pp0.07  & 2.44\pp0.02 &  1.49\pp0.31  &  0.60\pp0.10  & 0.90\pp0.03   &  0.07\pp0.05   \\  
	  &    Nuc    &  1.03\pp0.12  & 2.58\pp0.03 &  1.00\pp0.11  &  0.64\pp0.10  & 3.23\pp0.20   & 15.48\pp2.35   \\  
	  & (f)135\,N &  1.28\pp0.10  & 3.19\pp0.03 &  1.03\pp0.11  &  0.79\pp0.12  & 0.56\pp0.03   &  1.70\pp0.26   \\  
	  & (g)174\,N &  1.05\pp0.13  & 2.63\pp0.03 &  0.88\pp0.10  &  0.65\pp0.10  & 0.30\pp0.02   &  0.48\pp0.11   \\  
	  & (h)212\,N &  1.19\pp0.21  & 2.97\pp0.05 &  0.44\pp0.10  &  0.73\pp0.12  & 0.12\pp0.01   &  0.07\pp0.03   \\  
	  & (i)251\,N & ...   	      & ...	    &  0.45\pp0.14  &  ...	    & 0.01\pp 0.00  &  0.01\pp0.01   \\ 	 
	  & (j)290\,N & ...   	      & ...	    &  1.51\pp0.32  &  ...	    & ...	    &  0.03\pp0.02   \\ 	
	  & Mean      &  0.86\pp0.04  & 2.15\pp0.01 &  1.13\pp0.04  &  0.64\pp0.02  &   	   &        \\ 	
\noalign{\smallskip}
\hline
\noalign{\smallskip}
\end{tabular}
\begin{list}{Table Notes:}
\item (1) interstellar extinction; (2) visual extinction derived by the emission lines ratio; (3) visual extinction derived by {\sc starlight}; (4) color excess;
(5) SFR derived by the emission lines ratio; (6) SFR derived from {\sc starlight} output parameters. Whenever there are no error value, the uncertainties are zero.
\end{list}
\end{table*}

\subsection{Comparison between M05 and M11 models}

In order to deepen our analysis, we performed SP synthesis for the four galaxies using the new models computed by \citet[][hereafter M11]{maraston11}. 
These higher resolution models have been constructed in the same way as M05 (e.g. stellar energetics, the atmospheric parameters, the treatment of
the TP-AGB phase and horizontal branch morphology), but are based on different stellar spectral libraries (empirical libraries). The set of templates that
extend to the NIR are those based on the \citet{pick98} library. Rather than including spectra of individual stars, it practically averages out
spectra from several sources. However, since there are not enough spectra to cover all evolutionary phases to a tolerable degree for other than solar metallicities,
models were computed with Z=Z$_{\odot}$. Moreover, above $\sim$1$\mu$m about half of the spectra lack spectroscopic observations leading the authors
to construct a smooth energy distribution from broad-band photometry, resulting in a featureless region on the M11 spectra. 
This may imply that some NIR absorption features are not well resolved, even for these higher resolution models. 
The authors also call attention for the youngest ages (below $\sim$100\,Myr), which should be used somewhat more cautiously, once
their spectra are significantly attenuated by dust absorption in the UV.

\begin{figure*}
\includegraphics[width=0.9\linewidth]{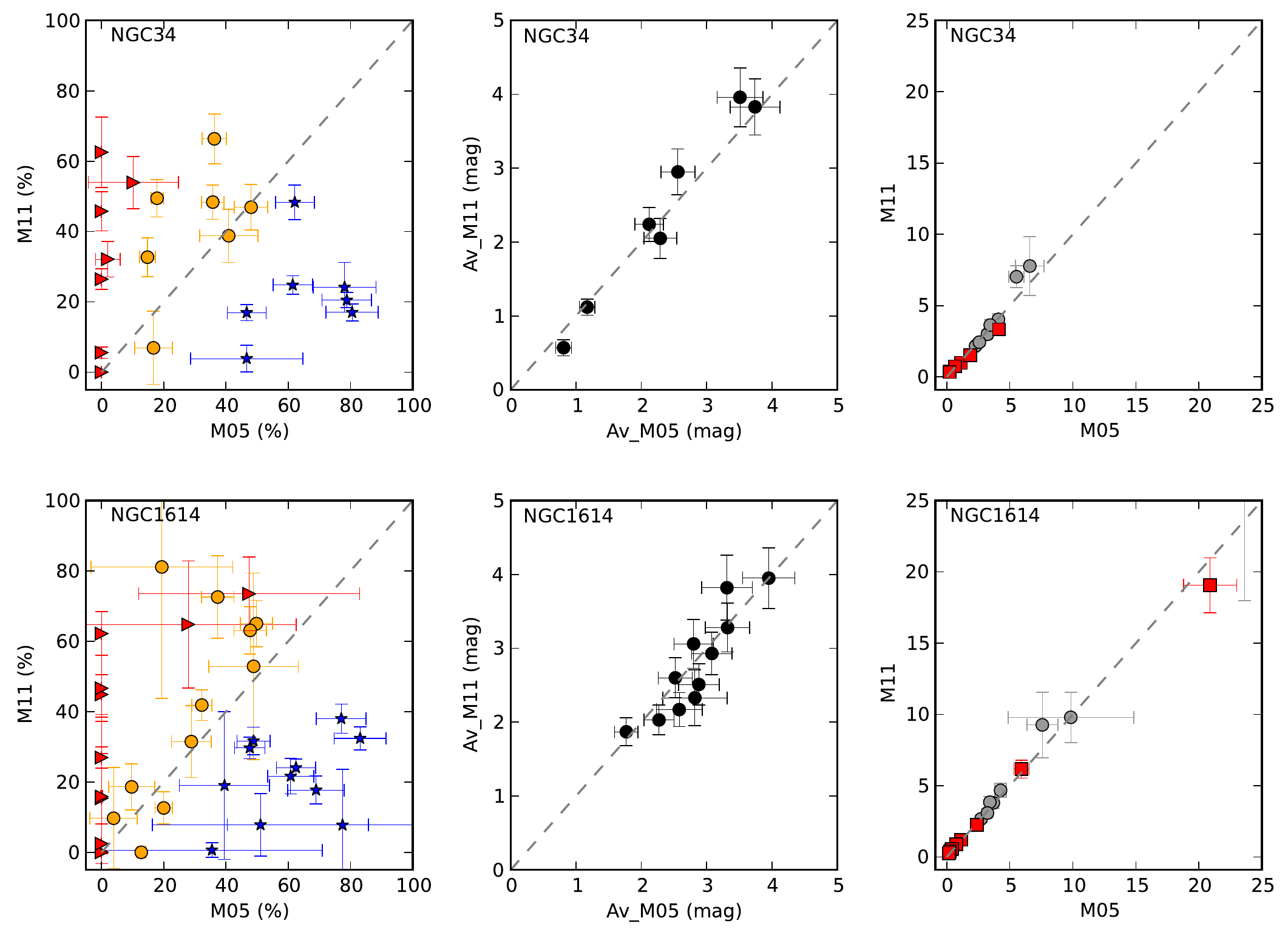}
\caption{Comparison between the results obtained using M05 and M11 models for galaxies NGC\,34 ({\it upper panels}) and NGC\,1614 ({\it bottom panels}). 
{\it Left Panels:} Stellar population distribution. The different markers indicate the stellar population 
age component:{\sc X$_y$} ({\it star}), {\sc X$_i$} ({\it filled circle}), {\sc X$_o$} ({\it triangle}). 
{\it Middle Panels:} Optical extinction ($A\rm_{v}$). {\it Right Panels:} Quality of the fit. The {\it square} marker
represents the {${\chi}^2$} and the {\it filled circle} the Adev.}
\label{compare1}
\end{figure*}

\begin{figure*}
\includegraphics[width=0.9\linewidth]{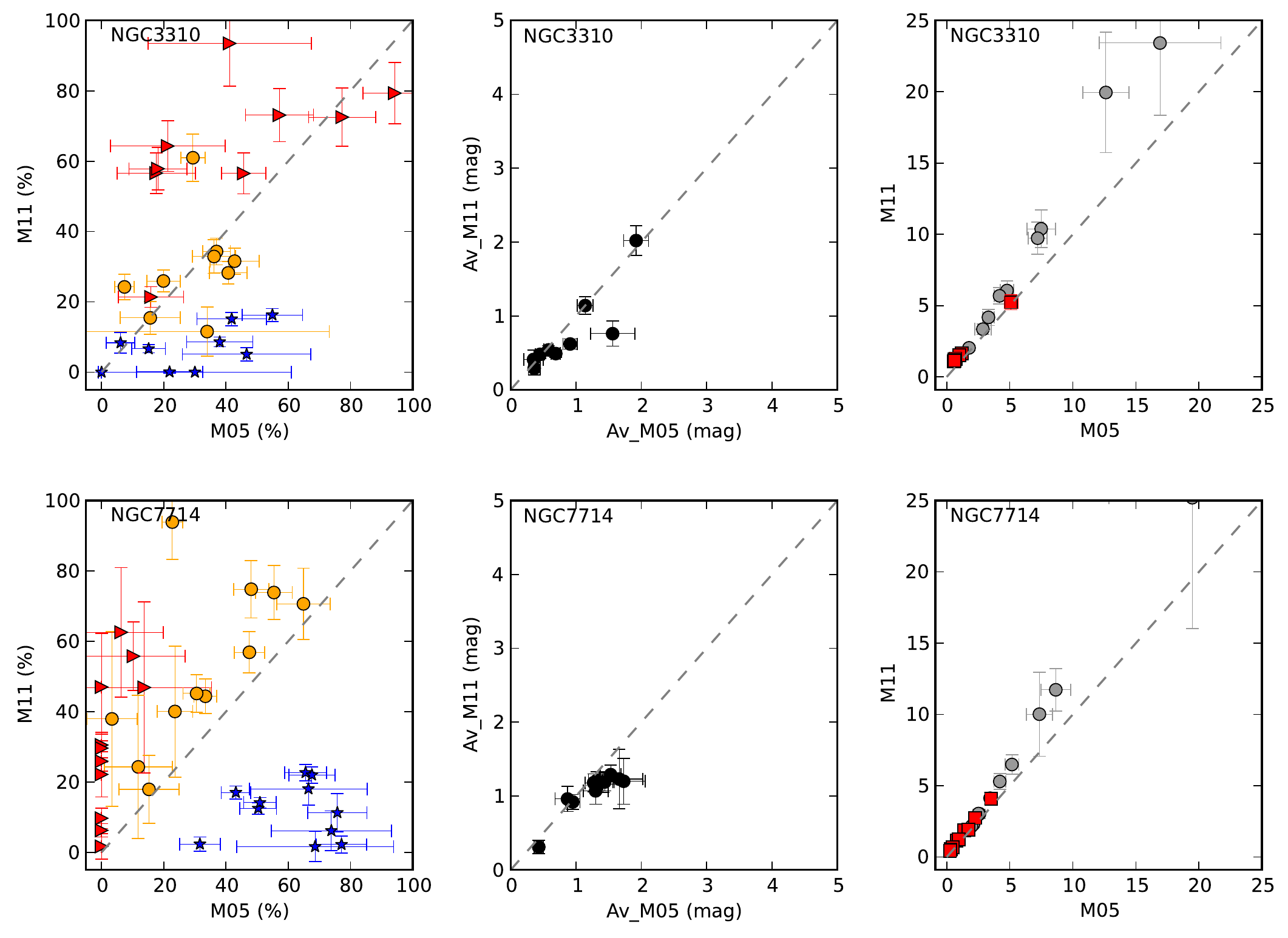}
\caption{Same as Fig.~\ref{compare1} but for galaxies NGC\,3310 ({\it upper panels}) and NGC\,7714 ({\it bottom panels}).}
\label{compare2}
\end{figure*}

To allow a proper comparison between both models sets, avoiding the discrepancies due to different metallicity ranges, we used a reduced base including only the solar metallicity 
SSPs from M05 models with the same age range used in section~\ref{resdisc}. Figs~\ref{compare1} and \ref{compare2} present the differences in {\sc starlight} output 
parameters {\sc X$_y$}, {\sc X$_i$}, {\sc X$_o$}, $A\rm_{v}$, Adev and ${\chi}^2$ between M05 and M11 models.

In general, M11 models enhance the old/intermediate age SP component in favor of younger ages when compared to M05, leading to small contributions of {\sc X$_y$} 
(the young component would not cross the limit of 40\% along the galaxies with the exception of the nucleus of NGC\,34 in which the young component reaches $\sim$48\%). 
This could be related with M11 models issue regarding younger ages, as mentioned before. The intermediate age SP component is the less affected by the choice of the model, 
however this SP component tends to get higher values with M11 models, mainly in two galaxies, NGC\,34 and NGC\,7714. These two sources are the ones that display higher 
young SP contribution when using M05 models with the full range in metallicities of our base set.  

The extinction $A\rm_{v}$ results are in good agreement (see Figs~\ref{compare1} and \ref{compare2} - {\it middle panels}). NGC\,3310 and NGC\,7714 though, 
tend to be less reddened when using M11 models, as well as they present higher Adev and ${\chi}^2$ values (indicators of the quality of the fit - see Sec.~\ref{spsm}), as we can see by analyzing the {\it middle} 
and {\it left panels} in Figs~\ref{compare1} and \ref{compare2}.

Due to the limitations of M11 models as discussed above, it is not clear whether the use of these higher resolution models would bring an improvement capable of
compensating the fact that they are only available for solar metallicities. Thus we conclude that the use of low-resolution M05 models are still the best option 
in this wavelength range.

\section{Conclusions}\label{conclusions}

We have studied, by means of NIR spectroscopy (from 0.8 to 2.4$\mu$m) the spatial variation of the
SPs in the central region of the local universe Starburst galaxies NGC\,34, NCG\,1614, 3310 and 7714. 
Therefore, we employed the {\sc starlight} code updated with M05 EPS models, which are the most suitable ones, once they
include a proper treatment of the TP-AGB phase crucial to model the stellar populations in the NIR \citep[M05, ][]{riffel11a}. 
Our main conclusions are:

\begin{itemize}
\item The near-infrared light dominating the off-nuclear apertures of the galaxies is due to young/intermediate age stars 
($t \leq 2\times10^9$\,yr), summing from $\sim$40\% up to 100\% of the light contribution. 

\item A predominance of young/intermediate age SP ($t \leq 2\times10^9$\,yr) is observed also in the central region of the galaxies, except for NGC\,1614 
in which the old SP ($t \geq 2\times10^9$\,yr) prevails in the nucleus, but the younger ages predominate the nuclear surroundings of this source. 

\item Evidence of a ring-like structure of about 600\,pc and a secondary nucleus at $\sim$300\,pc north from the nucleus was detected in NGC\,1614. 

\item The increase in the intermediate age SP component north from the nucleus in NGC\,7714 could be related to the SP present
in its companion, NGC\,7715.

\item The merger experienced by NGC\,1614, NGC\,3310 and NGC\,7714 can explain the discrepancy in its metallicity values, 
once the fresh unprocessed metal poorer gas from the destroyed/interacting companion galaxy is driven to the centre of the galaxies and mixed 
with the central region gas, before star formation takes place. In this context, the lower metallicity values derived for the young SP component ($t \leq 5\times10^6$\,yr)
can be understood as the diluted gas of the remnant.

\item As our sample present strong emission lines in their spectra we measure these nebular lines ([S{\sc iii}]\,$\lambda$9530, He{\sc i}\,$\lambda$10830, 
[Fe{\sc ii}]\,$\lambda$12570, Pa$\beta$\,$\lambda$12810, [Fe{\sc ii}]\,$\lambda$16440, H$_2$(1.0)S(1)$_o$\,$\lambda$21210 and Br$\gamma$\,$\lambda$21650) 
and derived values for the visual extinction ($A\rm_{v}$) and SFR along the galaxies. Our results on the interstellar extinction tend to agree with those 
found in the literature. We derived $\bar{A}\rm_{v}$=7.45\pp0.06, 2.94\pp0.01, 4.35\pp0.04 and 2.15\pp0.01 for NGC\,34, NGC\,1614, NGC\,3310 and NGC\,7714 respectively. 
Our values for the SFRs tend to be smaller than those calculated in other wavelengths. 

\item The comparison between M05 and M11 models indicates that M11 models tend to enhance the old/intermediate age SP contribution in favor of younger ages.
This could be related to the fact that the templates representing the youngest ages (below $\sim$100\,Myr) are significantly attenuated by 
dust absorption in the UV \citep{maraston11}. Moreover, above $\sim$1$\mu$m about half of the spectra lack spectroscopic observations leading the authors
to construct a smooth energy distribution from broad-band photometry. This may imply that some NIR absorption features are not well resolved, even for 
these higher resolution models. Due to these limitations of M11 models, it is not clear whether the use of these higher resolution models would bring 
an improvement capable of compensating the fact that they are only available for solar metallicities. Thus we conclude that the use of low-resolution 
M05 models are still the best option in the NIR wavelength range.

Therefore, the study of the distribution of the SP using the NIR spectral range are a useful tool in order to build a scenario for the star formation
along the sources. Moreover, the results for these star-forming interacting systems in the local universe can provide further support to the study of high-z sources,
which are of great importance for understanding the history of star formation in the early universe.

\end{itemize}

\section*{Acknowledgements}
We thank the anonymous referee for useful comments.
N.Z.D. and M.G.P. thank to CNPq for partial funding. R.R. is grateful to FAPERGs (ARD 11/1758-5), CNPq (304796/2011-5). A.R.A. acknowledges CNPq (307403/2012-2) for partial support to this work.
We all are deeply grateful to  Cid Fernandes and Charles Bonatto for the useful discussions. This research has made use of the NASA/IPAC Extragalactic Database (NED) which is operated by the Jet Propulsion 
Laboratory, California Institute of Technology, under contract with the National Aeronautics and Space Administration.


\appendix

\end{document}